\documentclass[aip,cha]{revtex4-1}

\usepackage{graphicx}
\usepackage{dcolumn}
\usepackage{amsmath}
\usepackage{amssymb}
\usepackage{color}
\usepackage{subfigure}

\begin{document}

\title{Connecting Lyapunov Vectors with the Pattern Dynamics of Chaotic Rayleigh-B\'enard Convection}

\author{R. Levanger}
\affiliation{Department of Electrical and Systems Engineering, University of Pennsylvania, Philadelphia, Pennsylvania 19104}
\affiliation{Department of Mathematics, Rutgers University, Piscataway, New Jersey 08854}

\author{M. Xu}
\affiliation{Department of Mechanical Engineering, Virginia Tech, Blacksburg, Virginia 24061}

\author{J. Cyranka}
\affiliation{Department of Computer Science and Engineering, University of California San Diego, La Jolla, California 92093}
\affiliation{Department of Mathematics, Rutgers University, Piscataway, New Jersey 08854}

\author{M. Schatz}
\affiliation{School of Physics, Georgia Tech, Atlanta, Georgia 30332}

\author{K. Mischaikow}
\affiliation{Department of Mathematics, Rutgers University, Piscataway, New Jersey 08854}

\author{M. R. Paul}
\affiliation{Department of Mechanical Engineering, Virginia Tech, Blacksburg, Virginia 24061}

\date{\today}

\begin{abstract}
We explore the chaotic dynamics of Rayleigh-B\'enard convection using large-scale, parallel numerical simulations for experimentally accessible conditions.  We quantify the connections between the spatiotemporal dynamics of the leading-order Lyapunov vector and different measures of the flow field pattern's topology and dynamics. We use a range of pattern diagnostics to describe the spatiotemporal features of the flow field structures which includes many of the traditional diagnostics used to describe convection as well as some diagnostics tailored to capture the dynamics of the patterns.  Using precision-recall curves, we quantify the complex relationship between the pattern diagnostics and the regions where the magnitude of the leading-order Lyapunov vector is significant.
\end{abstract}

\maketitle

\begin{quotation}
High-dimensional chaotic systems often yield striking patterns with intricate variations in space and time.  Examples include the dynamics of the weather, fluid turbulence, and reacting chemicals in a solution.  An important question to ask is: How do the structures of the patterns contribute to the overall disorder of the chaotic dynamics? We address this question using large-scale numerical simulations of chaotic Rayleigh-B\'enard convection for experimentally accessible conditions. Rayleigh-B\'enard convection is the buoyancy-driven motion that occurs when a shallow layer of fluid is heated from below. We also compute the growth and decay of small perturbations to the trajectory of the chaotic dynamics through state space by simultaneously computing the linearized dynamics.  We use the spatiotemporal dynamics of the leading-order vector describing these perturbations (the leading-order Lyapunov vector) to identify regions in the flow field patterns that are contributing significantly to the chaotic dynamics. We use a range of pattern diagnostics to quantify the connections between the convective pattern's topology and dynamics with the spatiotemporal dynamics of the leading-order Lyapunov vector.
\end{quotation}

\section{Introduction}
\label{section:introduction}

Many important challenges facing science and engineering today can be described as a spatially-extended system that is driven far-from-equilibrium~\cite{cross:1993}.  Examples include the complex dynamics of weather and climate, the patterns of interacting chemicals in biological systems, and the formation of uniform crystal structures from a melt.  In many cases of interest, the systems are strongly driven, the dynamics are highly nonlinear, and a model description (if available) is very complex and difficult to evaluate.  Furthermore, even if the model description is tractable, many of the important theoretical ideas and diagnostics are very difficult to compute.

One significant challenge is that the dimension of the dynamics of large, strongly driven systems is expected to be very large. For example, for many fluid systems under laboratory conditions, the dimension of the dynamics is expected to be on the order of hundreds or thousands~\cite{egolf:2000,duggleby:2010,karimi:2012}.  If the equations describing the dynamics are known, a powerful approach to probe the dynamics is through the computation of Lyapunov vectors and exponents~\cite{eckmann:1985,wolf:1985}.  Using current algorithms and computing resources, it is now possible to compute Lyapunov diagnostics for laboratory scale systems. 

However, in many systems that are beyond the current reach of numerics, it is possible to make detailed measurements of the dynamics.  This often yields large amounts of data describing the spatiotemporal patterns of quantities of interest.  For example, for a fluid system this could be many detailed images of the time variation of the flow field patterns. For a chemical system, this could be images of the spatiotemporal variation of the concentration of reactants and products. In geophysical problems, one could imagine using detailed satellite imagery of plankton blooms in the oceans or detailed weather dynamics as sources of data.

It would be very insightful to have a fundamental understanding of the connection between the pattern dynamics that can be measured in experiment and the deep insights that one gains from knowledge of the Lyapunov vectors and exponents.  In this article, we discuss our efforts to address this difficult problem using the canonical pattern-forming system of Rayleigh-B\'enard convection.

For several reasons, Rayleigh-B\'enard convection presents an opportunity to study the fundamental connections between powerful ideas from dynamical systems theory with experimentally accessible flow field dynamics.  Firstly, it is a pattern-forming system with a rich literature of deep insights from detailed experimental and theoretical studies~\cite{cross:1993,bodenschatz:2000}. Using modern algorithms and supercomputing resources, it is now possible to numerically simulate Rayleigh-B\'enard convection for the precise conditions of the experiment~\cite{paul:2003}. It is also possible to compute the Lyapunov vectors and exponents for the precise conditions of the experiment~\cite{scheel:2006,jayaraman:2006,paul:2007,karimi:2012}.

The article is organized as follows. In \S\ref{section:approach}, we briefly discuss Rayleigh-B\'enard convection, the Boussinesq equations, and our numerical approach for computing the flow field and the Lyapunov vectors and exponents. In \S\ref{section:results}, we describe our procedure for exploring the connection between the flow field topology and dynamics with the spatial variation of the leading-order Lyapunov vector.  We use the ideas of precision and recall to quantify the usefulness of a wide variety of pattern diagnostics in terms of their ability to indicate where the leading-order Lyapunov vector magnitude will be significant. Lastly, in \S\ref{section:conclusion} we present some concluding remarks.

\section{Approach}
\label{section:approach}

Rayleigh-B\'enard convection occurs when a shallow layer of fluid is uniformly heated from below in a gravitational field. As the temperature difference between the bottom and top surfaces is increased, the quiescent fluid layer undergoes a supercritical bifurcation to convective motion. This temperature difference can be represented as the Rayleigh number $R$ which often acts as the control parameter.  As the Rayleigh number is increased beyond its critical value $R > R_c$, the convection rolls themselves become dynamic.  Typically, the convection rolls become time dependent, chaotic, and eventually the rolls become unstable to smaller-scale structures, such as plumes, and the flow field becomes turbulent. 

Rayleigh-B\'enard convection contains all of the essential complexity and difficulties of spatially-extended systems driven far-from-equilibrium, yet it is experimentally accessible. With today's computing resources, it is also amenable to long-time numerical simulations for experimentally realistic conditions.

For moderate Rayleigh number, the Boussinesq equations describe the fluid motion. In nondimensional form, these are 
\begin{eqnarray}
\sigma^{-1}\left(\frac{\partial \vec{u}}{\partial t} + \vec{u} \cdot \nabla \vec{u} \right) &=&-\nabla p + \nabla^2 \vec{u} +RT\hat{z} \label{eq:momentum} \\
\frac{\partial T} {\partial t} + \vec{u} \cdot \nabla T &=& \nabla^2T \label{eq:energy} \\
\nabla \cdot \vec{u} &=& 0, \label{eq:mass}
\end{eqnarray}
which represent the conservation of momentum, energy, and mass, respectively. In our notation, $\vec{u}(x,y,z,t)$ is the velocity vector, $p(x,y,z,t)$ is the pressure, $T(x,y,z,t)$ is the temperature, $t$ is time, $(x,y,z)$ are Cartesian coordinates, and $\hat{z}$ is a unit vector in the $z$ direction which opposes gravity.  We have followed the typical convention and use the depth $d$ of the fluid layer as the length scale, the constant temperature difference $\Delta T$ between the bottom and top surfaces as the temperature scale, and the vertical thermal diffusion time for heat $d^2/\alpha$ as the time scale where $\alpha$ is the thermal diffusivity of the fluid.  The Prandtl number $\sigma$ is the ratio of the diffusivities of momentum and heat.  Finally, the extent of the cylindrical domain is given by the aspect ratio $\Gamma = r/d$, where $r$ is the radius of the domain.

The boundary conditions are no-slip at all surfaces in contact with the fluid.  The temperature of the hot bottom surface is $T(z\!=\!0)\!=\!1$, and the temperature of the cold top surface is $T(z\!=\!1)\!=\!0$. The lateral sidewalls of the domain are composed of a perfectly conducting material.

We also compute the spectrum of Lyapunov vectors and exponents. In the following, we provide only the essential ideas behind this computation and we refer the reader to Ref.~\cite{karimi:2012} for a more detailed discussion. In order to compute the $N_\lambda$ largest Lyapunov vectors, we simultaneously evolve $N_\lambda$ copies of the tangent space equations.  The tangent space equations are
\begin{eqnarray}
\sigma^{-1}\left(\frac{\partial}{\partial t} \delta \vec{u}^{(k)} + \vec{u}\cdot \nabla \delta \vec{u}^{(k)} +\delta \vec{u}^{(k)} \cdot \nabla \vec{u}\right)\qquad&=&-\nabla \delta p^{(k)} \label{eq:pert-momentum} \\ + \nabla^2 \delta \vec{u}^{(k)} +R\delta T^{(k)}\hat{z} \nonumber \\
\frac{\partial}{\partial t} \delta T^{(k)}+ \vec{u}\cdot \nabla \delta T^{(k)} +\delta\vec{u}^{(k)}\cdot \nabla T&=& \nabla^2 \delta T^{(k)} \label{eq:pert-energy} \\
\nabla \cdot \delta \vec{u}^{(k)}&=& 0 \label{eq:pert-mass}
\end{eqnarray}
where $k = 1, \ldots, N_\lambda$. The variables $\delta \vec{u}(x,y,z,t)^{(k)},\delta p(x,y,z,t)^{(k)}$, and $\delta T(x,y,z,t)^{(k)}$ are the $k^{\text{th}}$ perturbations about the nonlinear orbit through state space given by $\vec{u}(x,y,z,t)$, $p(x,y,z,t)$, and $T(x,y,z,t)$.

Using these perturbations, one can represent the $k^{\text{th}}$ Lyapunov vector at time $t$ as the large column vector given by
\begin{equation}
\vec{v}_g^{(k)}(t) = [\delta {u}(t)^{(k)}~\delta {v}(t)^{(k)}~\delta {w}(t)^{(k)}~\delta T(t)^{(k)}]^T,
\label{eq:vg}
\end{equation}
where $(\delta {u}^{(k)}, \delta {v}^{(k)}, \delta {w}^{(k)})$ are the $(x,y,z)$ components of the $k^\text{th}$ perturbation velocity field $\delta \vec{u}^k$, and the superscript $T$ indicates a transpose.  Equation~(\ref{eq:vg}) is meant to convey that the Lyapunov vector $\vec{v}^{(k)}_g(t)$ is represented as all of the values of $\delta u^{(k)}(t)$ at time $t$ as one large row vector, followed by all of the components of $\delta v^{(k)}(t)$ as one large row vector, and so on for $\delta w^{(k)}(t)$ and $\delta T^{(k)}(t)$. The perturbation of the pressure field $\delta p(t)^{(k)}$ is not included here since it is not an independent variable for incompressible flow.

In practice, the spectrum of Lyapunov vectors $\vec{v}_g^{(k)}$ are periodically reorthonormalized using a Gram-Schmidt procedure to avoid numerical errors associated with the tendency that all of the vectors will point toward the direction of fastest growth in the tangent space. The subscript $g$ on $\vec{v}_g^{(k)}$ indicates that these are the Gram-Schmidt orthogonalized vectors. As a result, the only Lyapunov vector pointing in a physically important direction will be the leading-order Lyapunov vector $\vec{v}_g^{(1)}(t)$.  However, the Gram-Schmidt reorthonormalization is volume preserving in the tangent space and, therefore, one can use these vectors to correctly compute the spectrum of Lyapunov exponents $\lambda_k$, which are guaranteed to be in the order $\lambda_1 \ge \lambda_2 \ge \ldots \lambda_{N_\lambda}$.

In order to numerically integrate Eqs.~(\ref{eq:momentum})-(\ref{eq:mass}) and the $N_\lambda$ copies of Eqs.~(\ref{eq:pert-momentum})-(\ref{eq:pert-mass}), we use a highly efficient parallel spectral element approach~\cite{fischer:1997} and its implementation in the open source solver  nek5000~\cite{nek5000}. The approach is third-order accurate in time, exponentially convergent in space, and has been used broadly to study a wide range of problems in fluid dynamics\cite{deville:2002}. A more detailed description of the numerical approach as it pertains to explorations of Rayleigh-B\'enard convection can be found in Refs.~\cite{paul:2003,jayaraman:2006,paul:2007,scheel:2006,karimi:2012}.

In this paper we focus our attention on the leading-order Lyapunov vector $\vec{v}_g^{(1)}(t)$ and its relationship to the flow field patterns and dynamics.  In order to probe this connection, we use the temperature perturbation field at the horizontal mid-plane $\delta T(x,y,z=1/2,t)^{(1)}$ as a representation of the Lyapunov vector~\cite{xu:2018}.  This allows us to identify the regions in physical space where the divergence of the Lyapunov vector is large and which we can connect with an experimentally accessible quantity such as the temperature field.

The approach we have taken is to explore ideas from pattern diagnostics and computational homology~\cite{kurtuldu:2011,kramar:2016} for use on the temperature field in order to build an understanding of their connection with the spatiotemporal variation of the leading-order Lyapunov vector.  This understanding provides physical insights into the complex dynamics of high-dimensional systems driven far-from-equilibrium. For example, it may lead to a description of which topological structures in the flow field are contributing most to the dynamics or perhaps a suggestion of what a modal decomposition of the dynamics may look like. These are difficult and open challenges. In this paper, we have pushed these ideas further for the canonical problem of Rayleigh-B\'enard convection. 

\section{Results and Discussion}
\label{section:results}

We have chosen to explore convection in a cylindrical domain with an aspect ratio of $\Gamma = 20.4$ containing a fluid with a Prandtl number $\sigma = 0.84$ and for a Rayleigh number of $R=4000$. These parameters are chosen in order to generate a high-dimensional chaotic state that is accessible to our numerical approach while also being in a regime that would be accessible to experiments using compressed gases~\cite{kramar:2016}.

The initial condition for our numerical simulation was a no-flow state with small, random perturbations to the temperature field. We then integrated Eqs.~(\ref{eq:momentum})-(\ref{eq:mass}) for a long time to allow all initial transients to decay.  This initial simulation was conducted for a time of $t = 800$ which is approximately equal to the two horizontal diffusion times $\tau_h$ for heat. The quantity $\tau_h$ is the amount of time it would take for heat to diffuse from the center of the domain to the sidewall, where $\tau_h = \Gamma^2$. The horizontal diffusion time for heat is expected to provide a useful estimate of the time required for initial transients to decay~\cite {cross:1984}.

\begin{figure}[tbh]
   \begin{center}
   \includegraphics[width=2.75in]{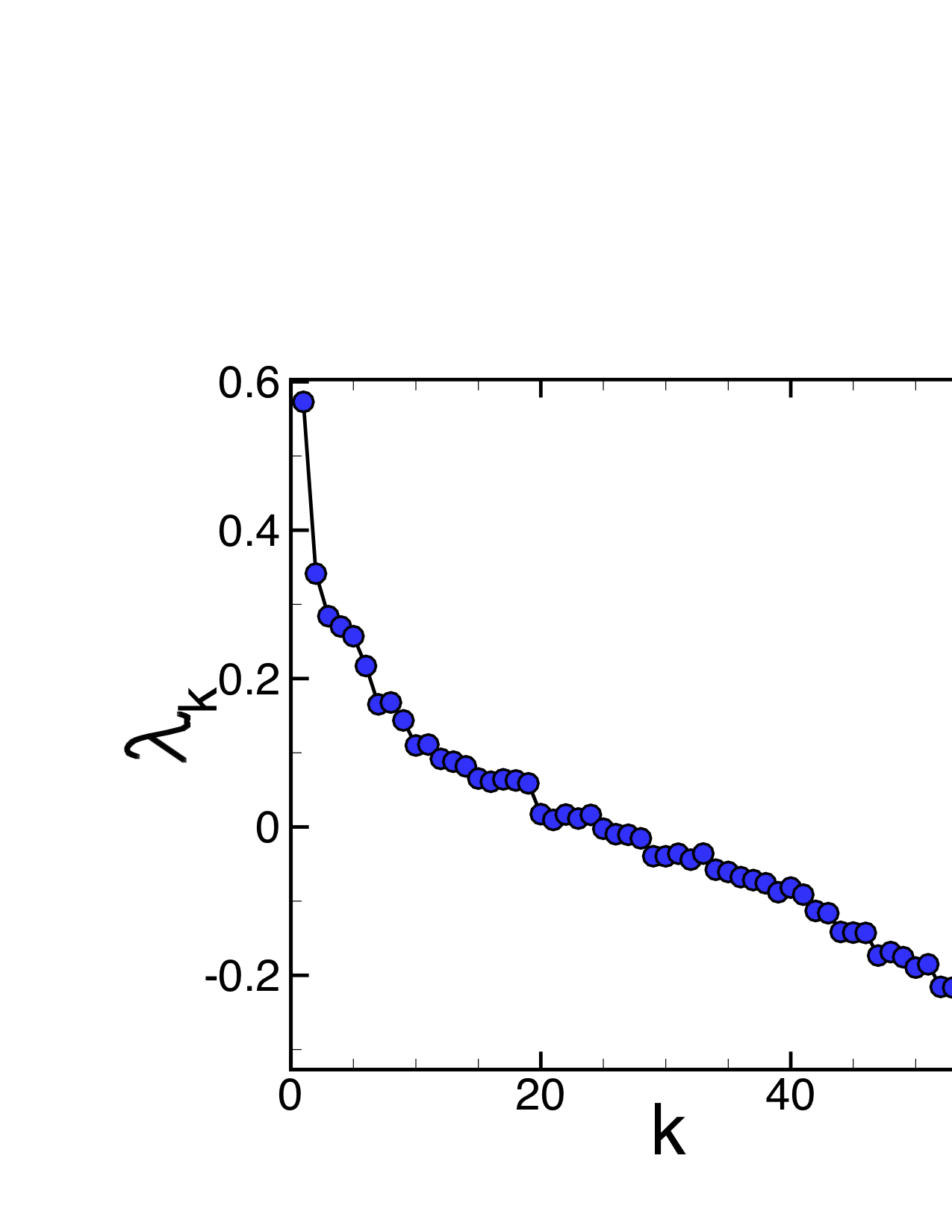}
   \includegraphics[width=2.75in]{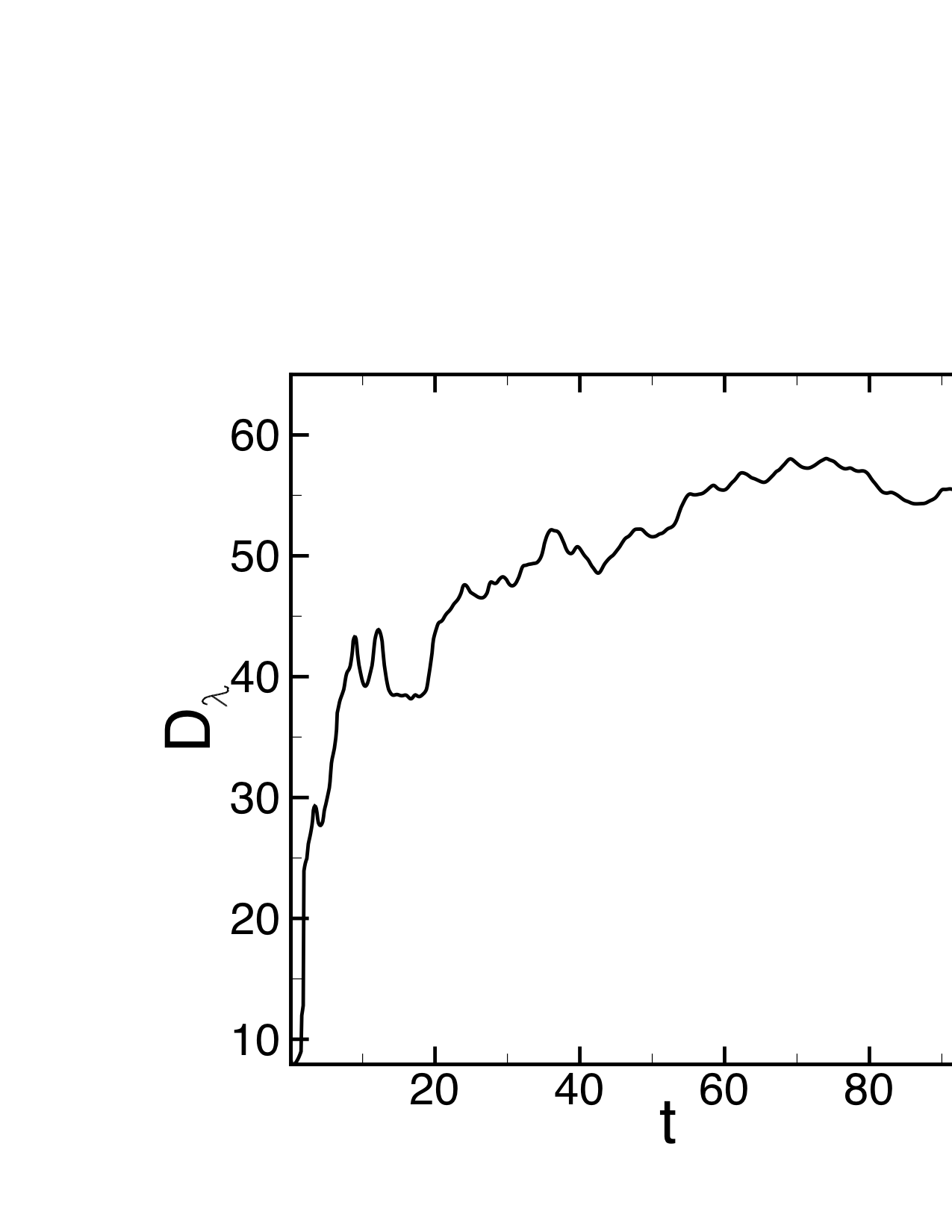}
   \caption{(a)~The spectrum of Lyapunov exponents $\lambda_k$ where $k = 1, 2, \ldots, N_\lambda$ is the index and $N_\lambda = 60$ is the number of Lyapunov exponents that were computed. (b) The time variation of the fractal dimension $D_\lambda$.  The long-time fractal dimension of the dynamics is $D_\lambda \approx 55$.}
    \label{fig:lyap}
    \end{center}
\end{figure}

We next numerically integrated Eqs.~(\ref{eq:momentum})-(\ref{eq:mass}) along with $N_\lambda = 60$ copies of Eqs.~(\ref{eq:pert-momentum})-(\ref{eq:pert-mass}) for another 100 time units. During this time, we performed periodic Gram-Schmidt reorthonormalizations and computed the Lyapunov vectors and exponents to generate the Lyapunov exponents and fractal dimension shown in Fig.~\ref{fig:lyap}.  We then continued the simulation for only the leading-order Lyapunov vector for another 350 time units.  The time step in our simulation was $\Delta t = 0.001$, and we performed the reorthonormalizations every 10 time steps. This yielded 3500 images of the flow field and leading-order Lyapunov vector that we have used in our analysis.

\begin{figure}[tbh]
   \begin{center}
   \subfigure[]{\includegraphics[scale=.525]{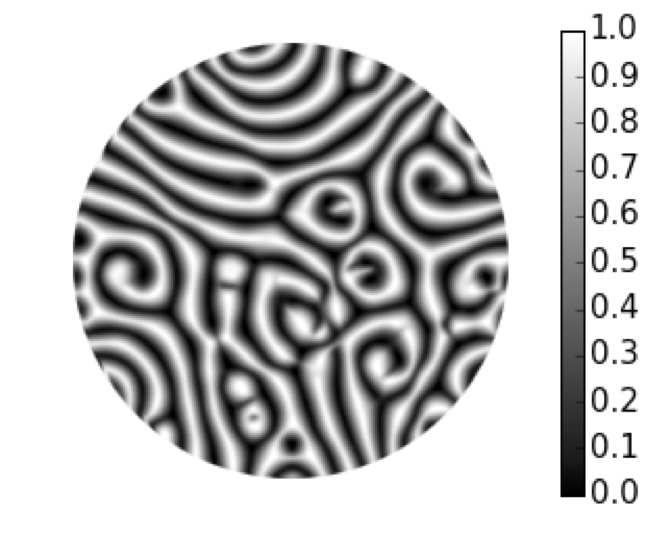}}
   \subfigure[]{\includegraphics[scale=.525]{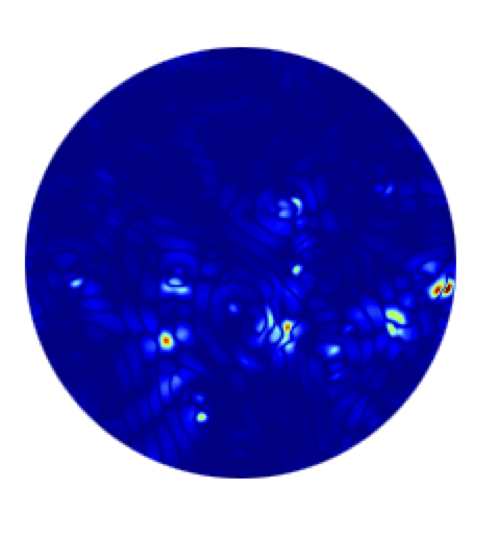}}
   \subfigure[]{\includegraphics[scale=.525]{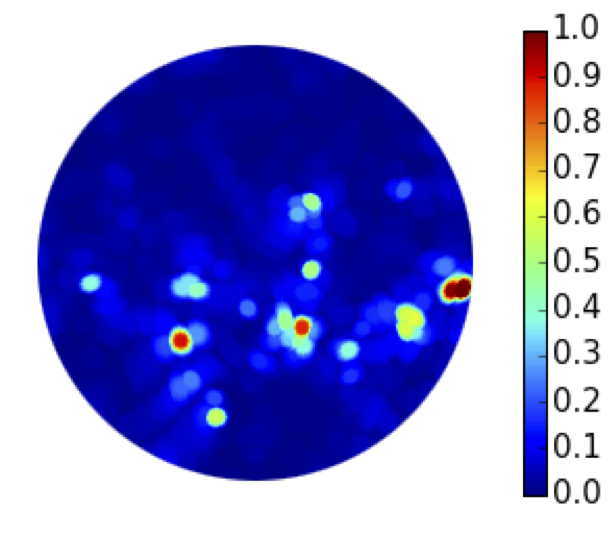}}
   \caption{(a)~An image of a typical flow field that is shown using the temperature of the fluid at the mid-plane $T(x,y,z=1/2,t)$, where light regions are hot rising fluid and dark regions are cool falling fluid. (b)~Color contours of the magnitude of the leading-order Lyapunov vector $|\delta T(x, y, z\!=\!1/2, t)^{(1)}|$ for the flow field shown in~(a) where red represents large values and blue represents small values. (c)~The image in (b), morphologically dilated by a disk of diameter of one unit length.}
    \label{fig:flowfield}
    \end{center}
\end{figure}

An image of a typical flow field pattern from our numerical simulation is shown in Fig.~\ref{fig:flowfield}(a). The image shows a horizontal midplane slice ($z\!=\!1/2$) of the temperature field, where light regions indicate hot rising fluid and dark regions indicate cool falling fluid.  The flow field pattern is composed of convection rolls and a variety of defect structures. Fig.~\ref{fig:flowfield}(b) shows the magnitude of the temperature perturbation field corresponding to the leading-order Lyapunov vector at the same horizontal midplane slice ($z\!=\!1/2$). We are interested in exploring the relationship between the high-magnitude regions of the leading-order Lyapunov vector and the topology and dynamics of the flow field pattern. However, one of our goals is to translate the results from this work from the numerical to experimental setting. Thus, the image shown in Fig.~\ref{fig:flowfield}(a) is derived by projecting the numerical results onto a standard 8-bit grey scale filtration that corresponds to an experimentally accessible image. To simplify the presentation, this grey scale filtration is itself rescaled to the unit interval. The use of the morphologically dilated leading-order Lyapunov vector magnitude function (Fig.~\ref{fig:flowfield}(c)) is explained in the next section.

Fig.~\ref{fig:lyap}(a) illustrates the spectrum of $N_\lambda=60$ Lyapunov exponents. The leading-order Lyapunov exponent is positive $\lambda_1 > 0$ which indicates that the dynamics are chaotic, as expected. Using the Kaplan-Yorke formula~\cite{kaplan:1979}, the fractal dimension $D_\lambda$ can be computed using only the Lyapunov exponents. Fig.~\ref{fig:lyap} illustrates the time variation of $D_\lambda$ as it approaches an asymptotic value of $D_\lambda \approx 55$. This indicates the dynamics has approximately 55 active chaotic degrees of freedom, on average.  Fig.~\ref{fig:lyap}(b) also illustrates the slow convergence of the fractal dimension.

\subsection{Evaluating diagnostic functions using precision and recall}
\label{section:precision-recall}

Recall that regions where the leading-order Lyapunov vector has high magnitude correspond to regions in the temperature field that exhibit a high degree of dynamic variability with respect to local perturbations. State-of-the-art methods for computing the leading-order Lyapunov vector apply only when one has full knowledge of the underlying dynamics, which we do not possess for experimental systems. Thus, our goal is to find experimentally accessible features that can be used to predict high-magnitude regions of the leading-order Lyapunov vector. For example, in Fig.~\ref{fig:flowfield}(b), we would like to be able to predict the spatial locations where the Lyapunov vector magnitude is significant (not dark blue) using only information from the pattern flow field that has been computed using images such as what is shown in Fig.~\ref{fig:flowfield}(a).  There are many possible  candidates to consider that may provide insight between the flow field patterns and the Lyapunov vector.  We have studied a variety of candidates that we separate into two groups.

In the first group, we consider many of the pattern diagnostics that are typically used when studying convection patterns. These are the spatial variation of the local wavenumber of the convection rolls and the presence of topological defects. Furthermore, we quantify the role of several particular types of topological defects including spirals, targets, grain boundaries, and disclinations.

In the second group, we include diagnostics that contain additional information that may be important indicators of where the Lyapunov vector is significant.  For example, the creation or annihilation of certain topological features as opposed to just their presence in the pattern. Specifically, we have explored temporal derivatives of the temperature field, the presence of roll pinch-off events, the emergence of target structures, and the dynamics of topological defects in the patterns.

Our approach to probe these different pattern diagnostics is the following. For each diagnostic, we define a pattern diagnostic function $F(x,y,t)$ that is based upon the temperature field at the horizontal mid-plane $z\!=\!1/2$ at time $t$. We use a threshold such that  $F(x,y,t)$ is a binary valued function where
\begin{equation}
F(x,y,t) =
\begin{cases}
1 & \text{diagnostic feature present} \\ 
0 & \text{diagnostic feature not present}.
\end{cases}
\end{equation}

We then seek to quantify the variation of $F(x,y,t)$ with the spatial variation of the leading-order Lyapunov vector at time $t$. Fig.~\ref{fig:flowfield}(b) illustrates that the spatial variation of the Lyapunov vector contains both localized and long-range structures. In a typical image of the leading-order Lyapunov vector, there are several localized maxima present which are indicated by the red regions which occur on a length scale of a single convection roll (approximately a nondimensional distance of unity).  However, there are also structures on a length scale of several convection rolls which are indicated by the light blue, green, and yellow regions. The interplay between these two length scales is dynamic and is not fully understood~\cite{xu:2018}.

In order to quantify the spatial variation of the leading-order Lyapunov vector we define another binary valued function $L(x,y,t;\alpha)$ where 
\begin{equation}
L(x,y,t;\alpha) =
\begin{cases}
1 & \text{if} \hspace{0.5cm}  |\delta T(x, y,t)^{(1)}| \ge \alpha \\ 
0 & \text{otherwise}.
\end{cases}
\end{equation}
The parameter $\alpha$ is a threshold that is used to determine how much of the Lyapunov vector we include in our analysis where $\alpha \in [0,1]$. When $x, y,$ and $t$ are understood, we will write $L(\alpha)$ instead of $L(x,y,t;\alpha)$. A large value of $\alpha$ would only include the localized peak structures in the Lyapunov vector. As the value of $\alpha$ is decreased, more of the larger-scale structure of the Lyapunov vector is also included. A value of $\alpha=0$ would simply include the entire Lyapunov vector field.

Having defined the binary valued functions $F(x,y,t)$ and $L(x,y,t;\alpha)$, we next quantify their relationship with one another which effectively reduces the problem to a binary classification task.  Since the proportion of the domain corresponding to high-magnitude regions of the Lyapunov vector is small (\emph{i.e.} imbalanced), we perform this comparison using precision-recall curves~\cite{Saito:2015,He:2009}, where the variational parameter is the threshold $\alpha$. Since we do not expect to have a perfect predictor $F$ for the Lyapunov vector magnitude function $L$, we also morphologically dilate $L$ by a disk with a diameter of one unit length which serves to increase the area accounted for by the local maxima~\cite{soille:2013} (see Fig.~\ref{fig:flowfield}(c)).

Consider a chaotic flow field at time $t$ where we have computed $F(x,y,t)$ and $L(x,y,t;\alpha)$.  The precision $\mathcal{P}_{\alpha, t}$ of the diagnostic function $F$ using a Lyapunov magnitude function $L$ at a threshold value of $\alpha$ at time $t$ is defined as the conditional probability given by
\begin{equation}
\mathcal{P}_{\alpha, t}(F,L) = \mathbb{P}\left[ L(x,y,t;\alpha) = 1 \mid F(x,y,t) = 1\right].
\end{equation}
In this definition, the probability $\mathbb{P}$ is computed using the values of $F$ and $L(\alpha)$ over all of the spatial points at time $t$. The precision $\mathcal{P}_{\alpha, t}$ is the probability that the Lyapunov magnitude function $L(\alpha)\!=\!1$ at some location in space given that the diagnostic indicator function $F\!=\!1$ at this location for a flow field at time $t$.

The precision of a diagnostic function is a measure of its predictive value. For example, consider a diagnostic function $F$ that identifies $N$ locations in a flow field image where $F\!=\!1$ while only $M$ of these locations also yield $L\!=\!1$ for a particular value of the threshold $\alpha$.  In this case, the precision of the diagnostic function is $\mathcal{P}_\alpha\!=\!M/N$. Of the regions identified by $F$, the precision is the probability that these identified regions correspond to a region where $L\!=\!1$.

Similarly, the recall $\mathcal{R}_{\alpha, t}$ of $F$ using $L$ at the threshold value $\alpha$ is defined as 
\begin{equation}
\mathcal{R}_{\alpha, t}(F, L) = \mathbb{P}\left[ F(x,y,t) = 1 \mid L(x,y,t;\alpha) = 1\right].
\end{equation}
The recall $\mathcal{R}_{\alpha, t}$ is the probability that the diagnostic indicator function yields $F\!=\!1$ given that the Lyapunov magnitude function $L(\alpha)=1$ at a particular point in space for a flow field at time $t$. Consider again our example where a diagnostic function $F$ has identified $N$ locations where $F\!=\!1$ in a flow field image where $M$ of these points also yield $L(\alpha)\!=\!1$.  In this case, we also use the fact that there are a total of $N_T$ locations identified such that $L(\alpha)\!=\!1$ where $N_T\!\ge\!M$. The recall of the diagnostic function is then $\mathcal{R}_\alpha\!=\!M/N_T$.  Therefore, the recall is the fraction of the total number of regions where $L(\alpha)\!=\!1$ that are identified by the diagnostic function $F$.

A large value of the recall indicates that a large percentage of the regions where $L(\alpha)\!=\!1$ are captured by the diagnostic function $F$. However, one could imagine a diagnostic where $F\!=\!1$ nearly everywhere on the pattern. This would yield a large value for the recall, yet such a diagnostic would clearly not be very useful. In practice, knowledge about both precision and recall of the diagnostic is what is insightful.

Ideally, one would like a diagnostic function with perfect precision and recall where $\mathcal{P}_{\alpha, t}(F,L)\!=\!\mathcal{R}_{\alpha, t}(F,L)\!=\!1$ for every time $t$. This would be achieved if regions where $F$ and $L(\alpha)$ are unity exactly coincide for a flow field image at time $t$ using a particular value of $\alpha$.  However, this is never achieved for the patterns we explore and, instead, we find a range of precision and recall values for a particular diagnostic function that vary as a function of $\alpha$. 

We are interested in using these ideas to quantify a statistical relationship between a particular pattern diagnostic and the magnitude of the leading-order Lyapunov vector.  In the following, we compute the variation of the precision and recall with the value of the threshold $\alpha$ for many flow field images from the time series described above.  We find it useful to compute the time average of the precision and recall over this time series which we denote as $\overline{\mathcal{P}}_\alpha(F, L)$ and $\overline{\mathcal{R}}_\alpha(F, L)$, respectively. For our data, we have a total of 3500 flow field images separated in time by 0.1 time unit. We compute the time averaged precision and recall over the range of threshold $\alpha \in [0,1]$ in increments of $\Delta \alpha\!=\!0.05$.

Since we are concerned with measuring how well a diagnostic predicts the regions of high magnitude in the Lyapunov vector, we do not include any time points $t$ in the average where that particular diagnostic returns $F\!=\!0$ for all spatial locations.  We compare the efficacy of each diagnostic by studying the average precision-recall curves over the sampled time series by plotting the variation of $\overline{\mathcal{P}}_\alpha$ and $\overline{\mathcal{R}}_\alpha$ with the threshold $\alpha$.

\subsection{Group I: The local wavenumber and topological defects}
\label{section:group-I}

The patterns of Rayleigh-B\'enard convection are composed of counter-rotating convection rolls and defect structures~\cite{cross:1993,chaikin:2000}. From a topological point of view, the simplest pattern is a field of straight-parallel convection rolls and any local deviation from this pattern is considered a topological defect.  

We have developed an automated approach to compute the pattern diagnostic functions for topological defects.  Our computational approach is briefly outlined as follows.  We first identify several types topological point defects present in the pattern using the algorithm of Bazen and Gerez~\cite{bazen:2002}.  In particular, we determine the presence of spiral structures, target structures, grain boundaries, dislocation pairs, convex disclinations, and concave disclinations. We accomplish this by clustering the computed topological point defects using single-linkage hierarchical clustering with a radius of $r\!=\!1$. These clusters of topological point defects are grouped together based upon the number of points in each cluster. For example, monopoles (singletons), dipoles (clusters with two points), \emph{etc.}, and then these groups are analyzed for further classification.

Monopoles are classified as either convex or concave disclinations depending on the charge of the point defect ($+1$ or $-1$, respectively). Dipoles with oppositely-charged point defects are classified as dislocations. A dipole with two positively-charged point defects is classified as either a target or a spiral.  In order to differentiate between targets and spirals we use persistent homology to detect the presence (or absence) of a local extremum that is situated between the point defects in the case of a target (or spiral)~\cite{kramar:2016}.  We classify clusters of point defects that are arranged in a near-linear sequence of alternating positively and negatively charged point defects as a grain boundary. Lastly, any cluster that does not fit into one of the above classification schemes we refer to as an unclassified topological defect, although there are some exceptions to this rule which we describe in more detail in Appendix A.
\begin{figure}[tbh]
   \begin{center}
   \subfigure[]{\includegraphics[scale=.5]{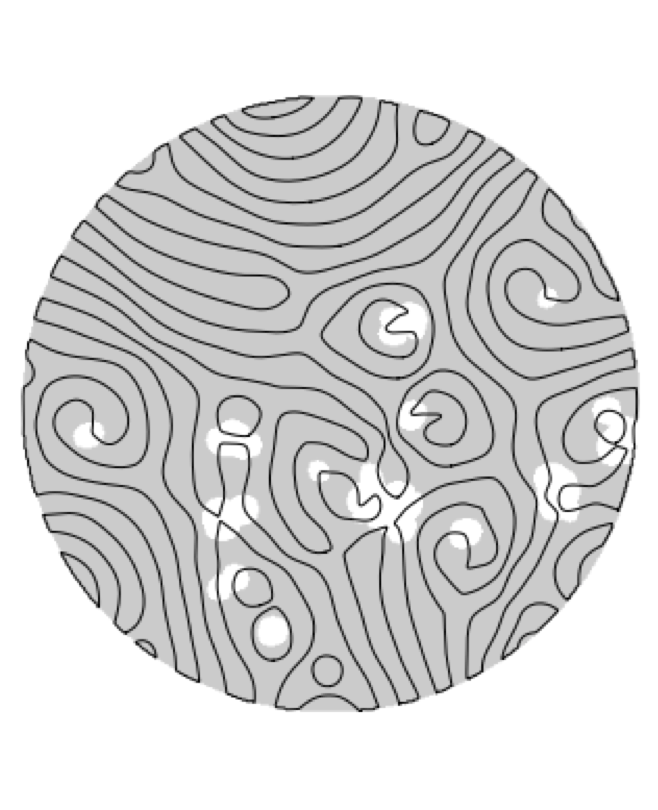}}
   \subfigure[]{\includegraphics[scale=.5]{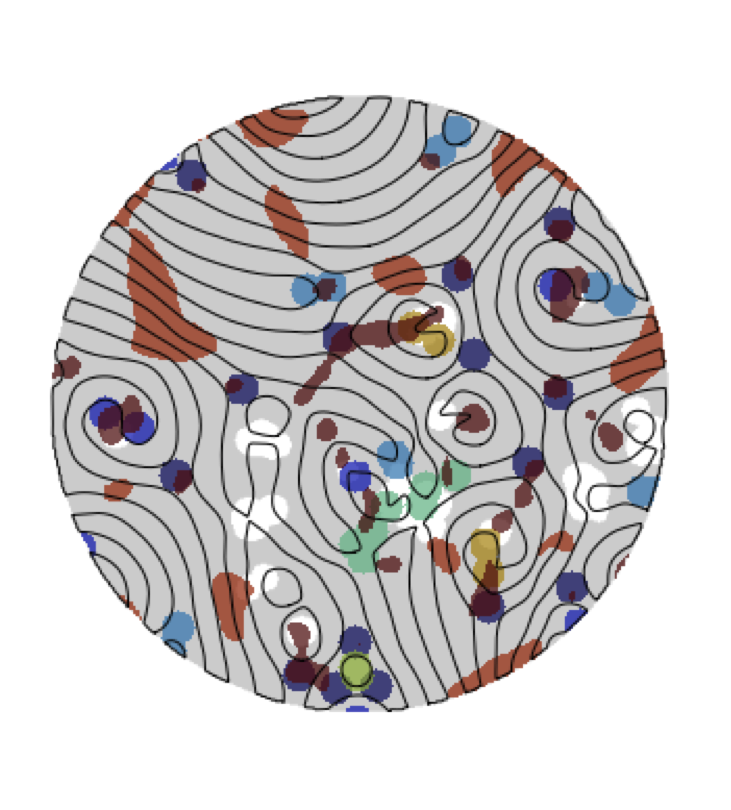}}
   \subfigure{\includegraphics[scale=.5]{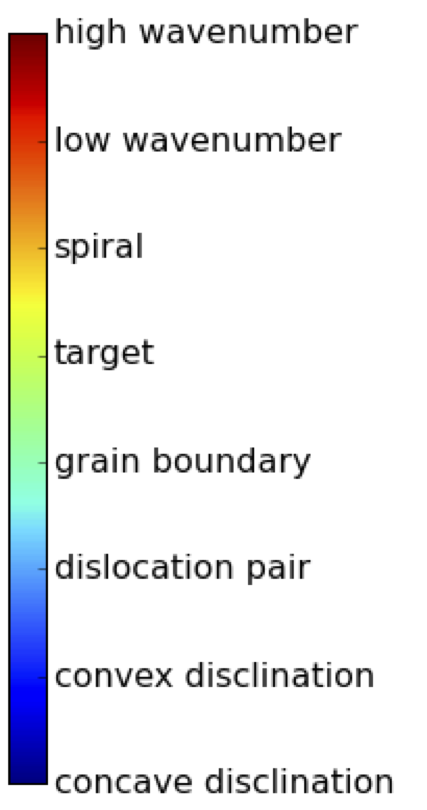}}
   \caption{A typical image of chaotic convection showing the complex relationship between eight different group I diagnostic functions $F$ and the Lyapunov magnitude function $L$ using a threshold of $\alpha\!=\!0.5$. In both (a) and(b), black solid lines indicate the pattern of convection rolls. (a-b) The Lyapunov magnitude function is represented as white wherever its value is above the threshold value $\alpha$ and as grey wherever it is below threshold. (b) Each different pattern diagnostic function is represented using a single color as indicated on the color bar. Spatial regions where the diagnostic function is beyond its threshold are indicated using that particular color. If the diagnostic function has not passed its threshold it is not represented at that point in space. The different pattern diagnostic functions represented are regions of high wavenumber, regions of low wavenumber, and the presence of a spiral, target, grain boundary, dislocation pair, convex disclination, and concave disclination.}
    \label{fig:group-I}
    \end{center}
\end{figure}

In addition, we use the local wavenumber $q$ of the pattern as a pattern diagnostic.  To compute the spatially varying local wavenumber we use the approach of Egolf \emph{et al.}~\cite{egolf:1998, becker:2006}. The wavenumber is an important parameter that is related to the linear stability of the convection rolls~\cite{cross:1993}.  The linear stability of an infinite field of straight and parallel convection rolls was computed in detail by Busse and co-workers~\cite{busse:1978}. The stability boundaries of straight and parallel convection rolls have proven to be insightful indicators of the pattern dynamics even for finite sized domains that are far from the critical threshold for convection. For example, when the local wavenumber of the pattern crosses a stability boundary this often results in a dynamic response by the pattern to return the wavenumber back within the bounds of linear stability.  As a result of this interplay between the pattern dynamics and well known results from linear stability, knowledge of the local wavenumber is an insightful quantity. 

For our investigation at constant values of $R=4000$ and $\sigma=0.84$, there are two particular linear stability boundaries of interest which occur at wavenumbers $q_+$ and $q_-$.  In our notation, $q_+$ is the stability boundary of straight and parallel convection rolls at large wavenumber, $q_-$ is the stability boundary at small wavenumber, and a local pattern wavenumber of $q$ where $q_- \!< \!q\! < q_+$ would be in the stable region.  A skew-varicose instability occurs for $q_+\! \approx \!2.7$ and a cross-roll instability occurs at $q_- \!\approx \!1.7$ (Ref. ~\cite{bodenschatz:2000}). We use these values of $q_+$ and $q_-$ as thresholds in defining two diagnostic functions. At spatial regions where the local pattern wavenumber $q\!>\!q_+$ we call this a \emph{high wavenumber} location and where $q\!<\!q_-$ we call this region a \emph{low wavenumber} location.

Fig.~\ref{fig:group-I} shows the pattern of convection rolls from Fig.~\ref{fig:flowfield} where we also use color contours to locate different diagnostic quantities of interest. The solid black lines represent the underlying pattern of convection rolls. The Lyapunov magnitude function $L(\alpha)$ is represented using the colors of grey and white where we have used a threshold value of $\alpha\!=\!0.5$ for this image. White indicates regions where the Lyapunov magnitude is above the threshold and grey indicates regions where it is below the threshold.  

Also shown on Fig.~\ref{fig:group-I} is the spatial variation of the eight different pattern diagnostic functions of group I where each particular diagnostic function is shown using a different color. The color bar indicates the particular color that is used to identify the location of each diagnostic function. If the diagnostic function is above its particular threshold, it is shown as its representative color at those regions. If the diagnostic function is below its particular threshold it is not shown on the figure.  The different diagnostic functions shown in Fig.~\ref{fig:group-I} represent regions of high local pattern wavenumber, regions of low local wavenumber and the presence of a spiral structure, target structure, grain boundary, a dislocation pair, a convex disclination, and a concave disclination. More details about the specific thresholds used and the methods for computing the classifications is located in Appendix A.

One important feature of the group I diagnostics is that one only needs an instantaneous image of the flow field pattern to compute these features. In other words, these diagnostics do not explicitly require information from previous times for their computation. We remove this restriction in \S\ref{section:group-I}, where we explore diagnostic functions that do require some knowledge of the flow field history for their computation.

Fig.~\ref{fig:group-I} conveys the complexity of the pattern and also the rich information contained in the group I pattern diagnostic functions. There are several interesting features present which are typical of these patterns and dynamics. First, the diagnostic functions highlight a diverse range of pattern features with some regions containing overlapping diagnostics with other regions only highlighted by a single diagnostic. The connection between the diagnostic regions and the regions where the Lyapunov magnitude function is above its threshold value (represented as white regions) appears quite complicated. For example, several white regions are highlighted by different diagnostic functions whereas several white regions are not highlighted by any of these diagnostics.

\subsection{Analysis of the group I diagnostics}
\label{section:analysis-group-I}

We next study the connection between the different group I diagnostic functions shown in Fig.~\ref{fig:group-I} and the Lyapunov magnitude function in detail. We accomplish this by quantifying 3500 images such as Fig.~\ref{fig:group-I} using the large set of data generated by our long-time numerical simulation of the chaotic dynamics.  We use the precision-recall approach described \S\ref{section:precision-recall} to generate a statistical description.

\begin{figure}[tbh]
   \begin{center}
   \includegraphics[width=\textwidth]{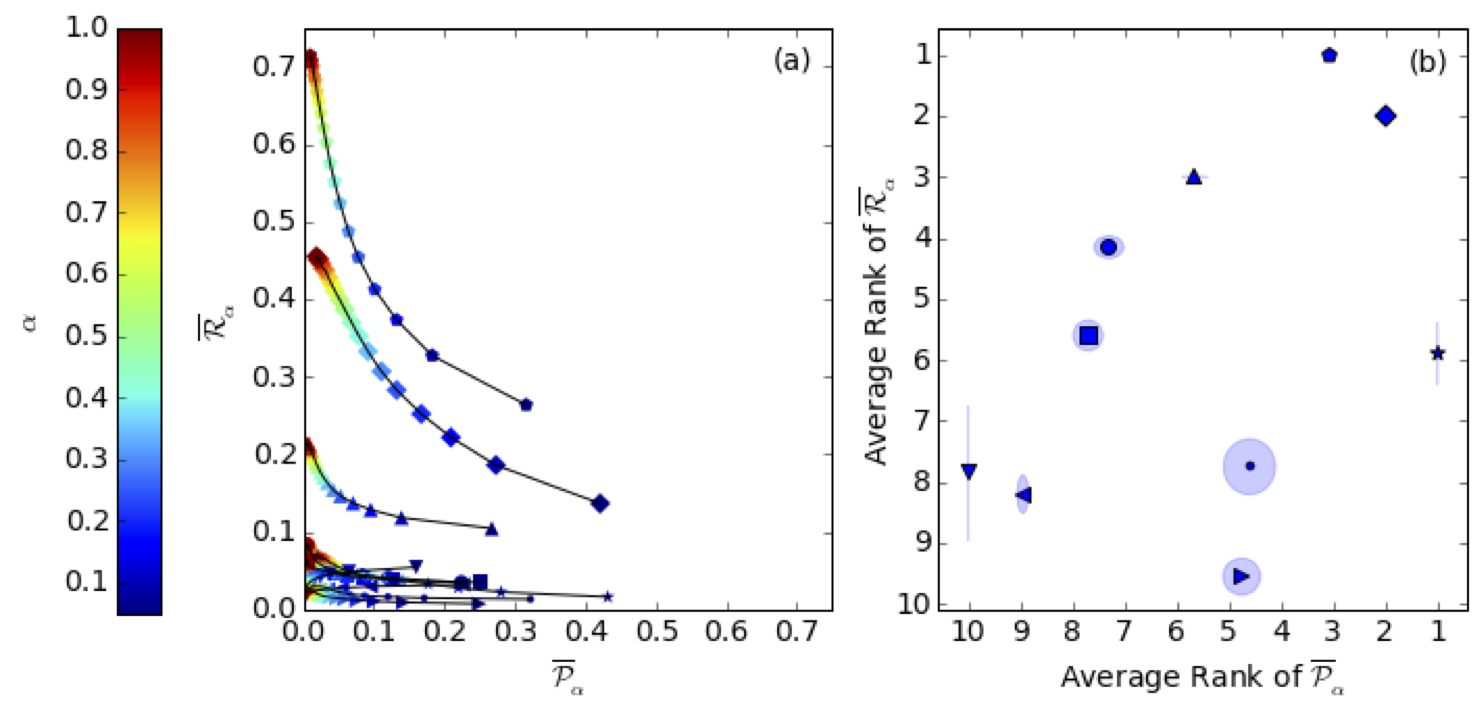}
   \caption{(a)~Average precision-recall curves for the group I pattern diagnostic functions. (b)~Average precision-recall rank over all values of the threshold $\alpha$ for the Lyapunov magnitude function. The light shaded region around the data symbols represents one standard deviation in the ranking distribution in the horizontal and vertical directions. The defects are indicated as follows: topological defects (pentagons), unclassified topological defects (diamonds), high wavenumber (up triangle), low wavenumber (down triangle), concave disclination (left triangle), convex disclination (right triangle), dislocation pair (circle), grain boundary (square), target (star), and spiral (dot).}
    \label{fig:prc-group-I}
    \end{center}
\end{figure}

Fig.~\ref{fig:prc-group-I}(a) shows the average precision-recall curves for the group I diagnostics. Each line represents the results for a different pattern diagnostic function as a function of the value of the threshold $\alpha$ used for the Lyapunov magnitude function. Each different pattern diagnostic function is also indicated using a different symbol. Each data point on a particular precision-recall curve is the average precision and average recall for that diagnostic using that value of $\alpha$ where the different values of $\alpha$ are indicated in color.

In addition to the eight diagnostics for specific types of topological defects we also include two new aggregate classifications in  Fig.~\ref{fig:prc-group-I}. We group together \emph{all} of the topological defects in one diagnostic function which we refer to as the topological defect diagnostic function. In addition, in the process of analyzing the patterns there were topological point defects identified which did not fit the signature of any of the particular defect structures that we have included. Any topological point defect that was not previously classified as a particular type of defect we have gathered together into the pattern diagnostic function that we refer to as unclassified. 

The precision-recall curves shown in Fig.~\ref{fig:prc-group-I}(a) exhibit some general trends. As the threshold $\alpha$ is reduced the recall decreases while the precision increases. For a large value of $\alpha$, such as $\alpha \approx 0.9$, this would indicate that only the spatial regions where the largest peaks of the Lyapunov magnitude function occur are included by setting $L(\alpha)=1$ at these locations. As a result, there are fewer regions highlighted by $L(\alpha)$ and these regions will be located near the largest peaks of the Lyapunov magnitude function. The average precision $\overline{\mathcal{P}}_\alpha$ of the different diagnostic functions will be low since they typically select numerous regions in the pattern yet now only a few regions have $L(\alpha)=1$. The result is that the pattern diagnostic functions will select regions that do not have $L(\alpha)=1$ which results in a lower precision. In addition, the average recall $\overline{\mathcal{R}}_\alpha$ will be large since the pattern diagnostic functions will successfully locate more of the regions where $L(\alpha)=1$, on average.

Conversely, it is useful to also describe the results when the threshold $\alpha$ is small, such as $\alpha \approx 0.1$. In this case, since the threshold for $L(\alpha)$ is small, there are many regions in space where the Lyapunov magnitude function is larger than the threshold. As a result, the precision of the diagnostic functions will increase. This can be understood as follows: Of the locations identified where the pattern diagnostic function yields $F=1$ there will now be more locations where $L(\alpha)=1$ which will yield a larger precision. However, for the smaller value of the threshold $\alpha$ the recall will be smaller. This is because at the lower value of $\alpha$ there are now many locations where $L(\alpha)=1$ which will not also be captured by the pattern diagnostic function.

The only diagnostic that does not follow these trends are the regions where the local wavenumber is below the threshold. However, this diagnostic does not perform well for either precision or recall and we have not investigated this diagnostic further.

It is interesting to highlight that the pattern diagnostic function for target structures yields a large amount of precision although its recall is quite small. This indicates that when a target structure is present in the flow field there is a significant probability that it will also be in a region where the leading-order Lyapunov magnitude function is above its threshold. However, there are typically very few target structures present in the pattern at any one time which results in a low value for the recall.

Of all of the group I diagnostics represented in Fig.~\ref{fig:prc-group-I}(a), there are three which stand out in terms of achieving a significant amount of recall. These are the aggregate of all of the topological defects, the aggregate of all of the unclassified topological defects, and the regions where the local wavenumber is large. It is interesting that the individual topological defects we have quantified do not perform well yet, on average, the aggregate of all of the topological defects performs well. This suggests that there is not a particular topological defect structure, that we have identified in group I, that is a good indicator of where the Lyapunov magnitude function will be above threshold. Rather, this suggests that the different diagnostics capture features of the Lyapunov vector that, when added together, performs better.

In addition, the aggregate of all of the unclassified topological defects performs well in terms of both precision and recall.  It is notable that the aggregate of the unclassified topological defects has a higher precision than the aggregate of all of the topological defects.  This indicates that removing the specific topological defects we have identified from the rest of the topological defects selects specifically for regions that have a higher correspondence with the high-magnitude regions of the Lyapunov vector.  This again suggests that the presence of a topological defect is what is important and not its classification as one of the typical canonical defects.

Lastly, the regions where the wavenumber is large also yields significant precision and recall. The connection between the Lyapunov vector and regions of large wavenumber has been discussed in the literature~\cite{egolf:1998,scheel:2006,paul:2007,karimi:2012}. However, it is interesting to point out that the local wavenumber is outperformed as a diagnostic, in terms of both precision and recall, by the pattern diagnostic functions for the aggregate of all topological defects and for all unclassified topological defects.

Using the data presented in Fig.~\ref{fig:prc-group-I}(a) we have computed the average precision and average recall of each diagnostic over all of the threshold values $\alpha$. Using these average values we rank each of the diagnostics with respect to one another in terms of precision and recall which is shown in Fig.~\ref{fig:prc-group-I}(b).  The light shaded region around each symbol indicates the standard deviation about the mean. This allows one to discern between the different diagnostics and to compare their performance with one another on average.  

The top three diagnostics in terms of the average recall are the aggregate of all topological defects, the aggregate of all of the unclassified topological defects, and the regions of large wavenumber.  In particular, the aggregate of all of the topological defects is ranked first in average recall and is ranked third in average precision. It is also apparent from Fig.~\ref{fig:prc-group-I}(b) that while the diagnostic function for target structures ranks first in precision it ranks fifth in terms of recall.

\subsection{Group II diagnostic functions}
\label{section:group-II}

The leading Lyapunov vector quantifies the growth of perturbations in the tangent space and, therefore by definition, explicitly describes changes in the system over time.  By contrast, the group I diagnostics, which are computed at a given instant on a single flow field image, can capture time-dependence only in an indirect, implicit manner.  For example, when the local wavenumber pattern diagnostic is computed on a pattern snapshot, large or small values of $q$ in the pattern suggest that stability boundaries have been crossed and, as a consequence, spatially-localized, instability-driven time-dependence is imminent.    In this section,  we examine pattern diagnostic functions that include time-dependence explicitly with the goal of determining whether such diagnostics are more effective in locating regions of high magnitude of the leading Lyapunov vector.   These diagnostics, which examine short sequences of pattern snapshots from flow image time series, probe topological point defects with a large velocity, the creation and annihilation of topological point defects, and temporal derivatives of the temperature field of the fluid at the horizontal midplane.   Additionally, in this section, we introduce two pattern diagnostics for single flow field images that identify, in a novel way, 
 roll pinch-off events and the emergence of target structures.

\begin{figure}[tbh]
   \begin{center}
   \subfigure[]{\includegraphics[scale=.5]{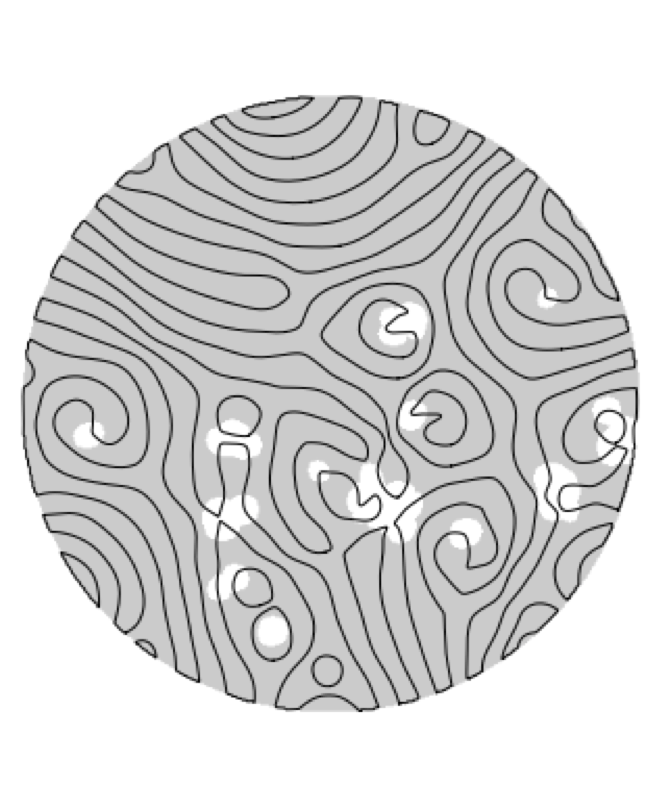}}
   \subfigure[]{\includegraphics[scale=.5]{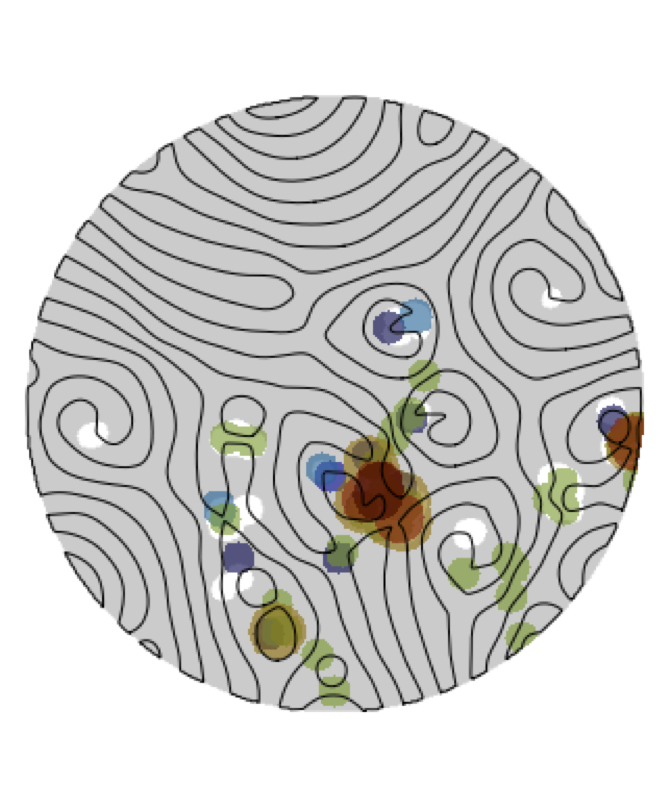}}
   \subfigure{\includegraphics[scale=.5]{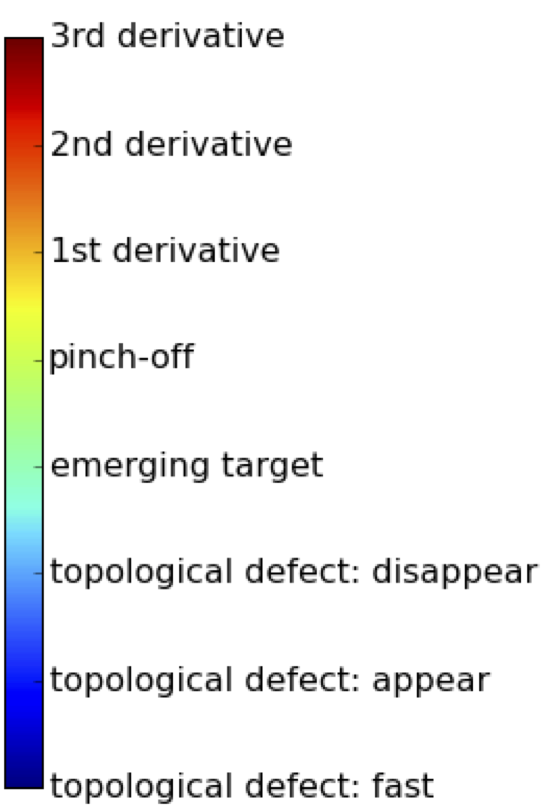}}
   \caption{An image of chaotic convection that also shows the group II diagnostic functions and the Lyapunov vector magnitude function using a threshold value of $\alpha=0.5$.  In both (a) and(b), the solid black lines indicate the underlying pattern of convection rolls. (a-b) Regions where the Lyapunov magnitude function is above the threshold $\alpha$ are shown in white and regions where it is below the threshold are grey.  (b) Each different diagnostic function is represented using a different color. The diagnostics included are the third-order, second-order, and first-order time-derivatives of the temperature, roll pinch-off events, emerging target structures, the annihilation of a topological defect, the creation of a topological defect, and topological defects with a large velocity (labeled `topological defect: fast').}
    \label{fig:group-II}
    \end{center}
\end{figure}

Fig.~\ref{fig:group-II} illustrates the relationship between a chaotic flow field image, the Lyapunov magnitude function, and the group II diagnostic functions at a particular instant of time. The solid black lines indicate the pattern of convection rolls. The Lyapunov magnitude function is shown using a threshold of $\alpha=0.5$ where white indicates regions above threshold and grey indicates regions below threshold.  Each pattern diagnostic is represented using a different color as indicated in the color bar. Fig.~\ref{fig:group-II} is using the same flow field image as that of Fig.~\ref{fig:group-I} which allows for a direct comparison between the identified group I and II diagnostic functions for this flow field image.

A roll pinch-off event can be identified as a saddle point in the temperature field at the horizontal midplane. Persistent homology can be used to locate all of the saddle points in the flow field image. However, in a typical convection flow field there are many saddle points located throughout the pattern due to the complex sinusoidal variations of the pattern of convection rolls.  The saddle points that are not associated with the roll pinch-off events can be filtered out.  The spurious saddle nodes will not be contained in a range of values of the temperature field. We have found that saddle nodes that occur outside of the range $0.2 \le T \le 0.8$ are spurious, and those that occur within this range are roll pinch-off events.
\begin{figure}[tbh]
   \begin{center}
      \includegraphics[width=0.65\textwidth]{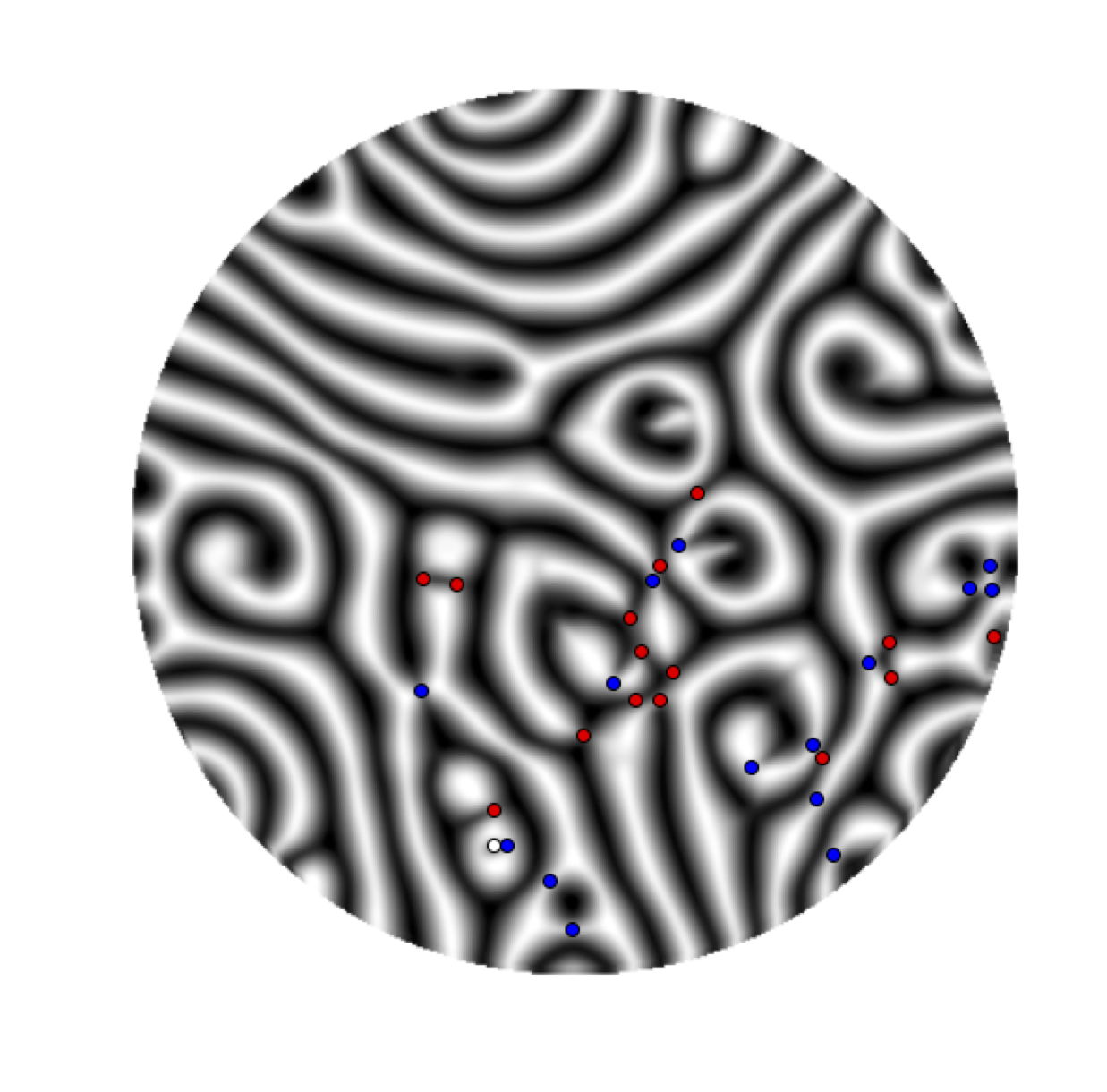}
   \caption{Grey-scale image of the temperature field at the horizontal mid-plane illustrating a pattern of convection rolls.  Red and blue circles indicate the presence of a saddle point and the white circle indicates the presence of a local minimum. All saddle points and local extrema that are selected are in the range $0.2 \le T \le 0.8$.}
    \label{fig:defects}
    \end{center}
\end{figure}

Emerging targets are characterized by the existence of a local extremum that has a value of the temperature in the range $0.2 \le T \le 0.8$.
We use persistent homology to locate all local extrema in the temperature field and then, as with saddle points, filter their values. For reasons associated to discretization errors, we filter out any saddle points and local extrema that are tied to persistence points with a lifespan less than $\delta T = 0.1$. The lifespan is a measure of the vertical distance a persistence point is from the diagonal in a persistence diagram.

Fig.~\ref{fig:defects} illustrates the use of persistence homology to identify the roll pinch-off events (blue and red circles) and emerging targets (white circles). Both of these defects can be seen as indicating the presence of a local minimum in the amplitude of the pattern. Fig.~\ref{fig:defects} shows the chaotic flow field using a grey scale to indicate the spatial variation of the temperature. The red and blue circles indicate the location of a saddle point and the white circle indicates a local minimum. In both cases, the values of the selected points occur in the range $0.2 \le T \le 0.8$.
\begin{figure}[tbh]
   \begin{center}
      \includegraphics[width=0.8\textwidth]{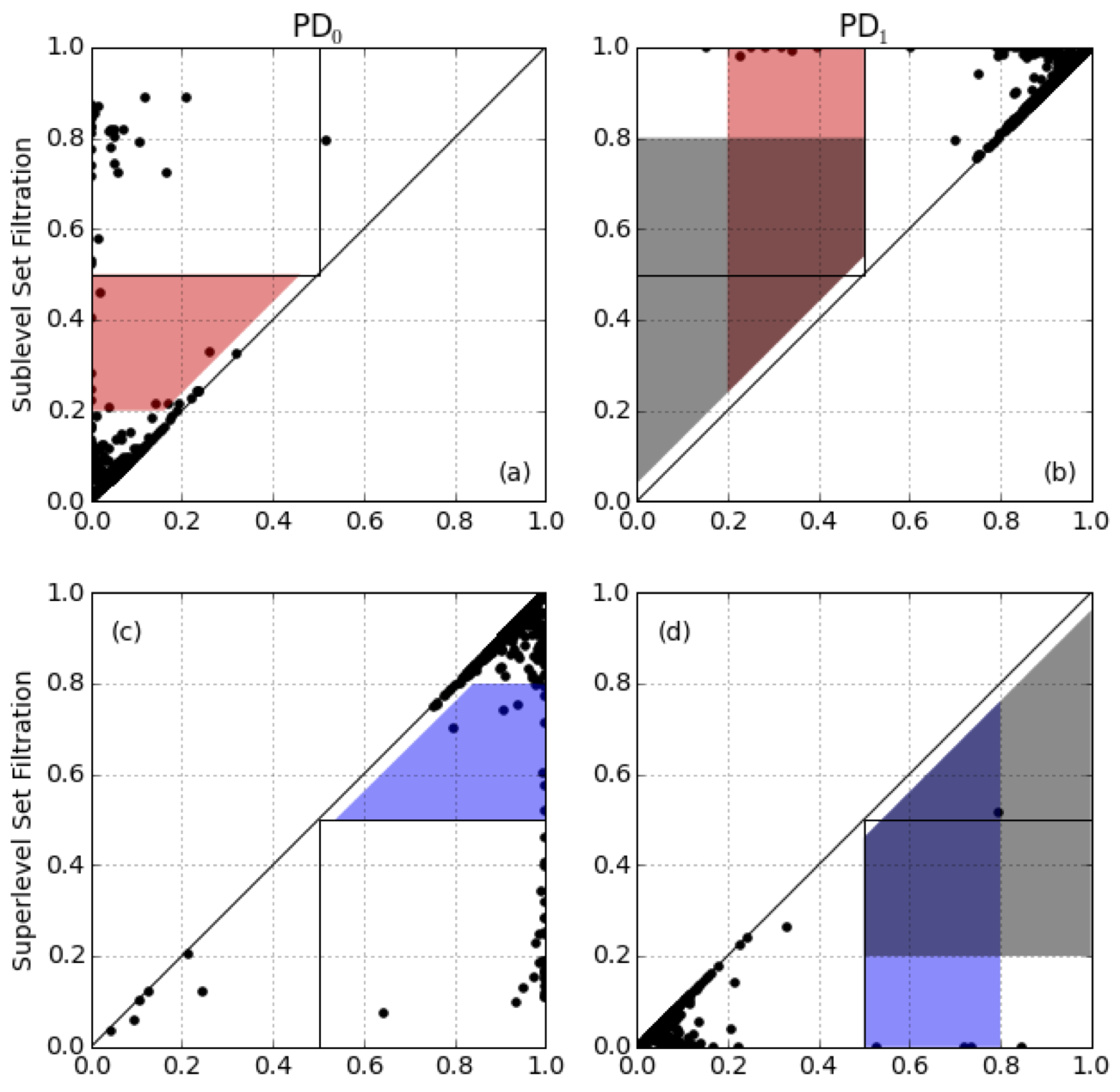}
\caption{The persistence diagrams used to compute the presence of roll pinch-offs and emerging target structures for the flow field image shown in Fig.~\ref{fig:flowfield}. Each black point in these figures is a persistence point. The horizontal and vertical axes in all of the plots are the temperature $T$. The top row (a)-(b) contains two persistence diagrams determined using sublevel sets and the bottom row (c)-(d) contains two persistence diagrams computed using superlevel sets. The red and blue regions indicate the partitions used in the persistence plane for both sublevel and superlevel sets to find the pinch-offs (saddle points). The grey regions indicate the filtration used to find emerging target structures (local extrema).} 
    \label{fig:persistence}
    \end{center}
\end{figure}

Fig.~\ref{fig:persistence} shows the corresponding regions in the persistence diagrams that are used to select the saddle points and local extrema for this annotation. A detailed description of the use of persistence diagrams for convection flow fields can be found in Ref.~\cite{kramar:2016}. For our purposes, we use the fact that each persistence point on a persistence diagram corresponds to a pair of critical points (saddle points or local extrema) in the underlying temperature field. The relationship between the type and dimension of the persistence diagram and the types of paired critical points is summarized in Table~\ref{tab:table1}. Thus, by selecting regions on each persistence plane, we are able to select particular critical points in the temperature field. More precisely, the red regions are used to locate saddle points corresponding to pinch-off events occurring in lower temperature ranges ($0.2 \leq T \leq 0.5$), the blue regions are used to locate saddle points corresponding to pinch-off events occurring in higher temperature ranges ($0.5 \leq T \leq 0.8$), and the grey regions are used to locate local extrema corresponding to emerging target structures. In all cases, persistence points that are a vertical distance of less than 0.04 temperature units from the diagonal are not selected. This ensures that only saddle points and local extrema associated to larger dynamical structures are selected.

\begin{table}[h!]
  \begin{center}
    \begin{tabular}{l|c|c|c} 
      \textbf{Filtration Type} & \textbf{Persistence diagram} & \textbf{Lower critical value} & \textbf{Higher critical value}\\
      \hline
      Sublevel set & PD$_0$ & Local min. & Saddle pt. \\
      Sublevel set & PD$_1$ & Saddle pt. & Local max. \\
      Superlevel set & PD$_0$ & Saddle pt. & Local max. \\
      Superlevel set & PD$_1$ & Local min. & Saddle pt.
    \end{tabular}
    \caption{Critical point (saddle points or local extrema) pairings underlying the persistence points on the persistence diagrams associated to a scalar field with a two-dimensional domain (see Fig.~\ref{fig:persistence}).}
    \label{tab:table1}
  \end{center}
\end{table}

We have also quantified the dynamics of the topological defects. In particular, we have used pattern diagnostic functions to locate regions containing  topological defects with a large velocity, to locate regions where topological defects are created, and to locate regions where topological defects are annihilated. This is done by matching topological point defects with the same charge from one frame in time to the next (see Appendix B for details). We then use these results to determine which topological defects that have been created, annihilated, or moved farther than a threshold value of $d = 0.2$ where the duration between frames is 0.1 time units.

As a measure to identify regions where the pattern changes are rapid in time we have also computed the temporal derivative of the temperature field at the horizontal midplane $\partial^n T/\partial t^n$ where $T(x,y,z\!=\!1/2,t)$ and $n$ is the order of the derivative. We have explored $n=1,2,3$ using first-order-accurate finite differencing to compute the temporal derivatives of the temperature field. To determine if the temporal derivative is changing rapidly we have used the threshold $\partial^n T/\partial t^n \ge 0.16$, which was determined using trial and error.

\subsection{Analysis of the group II diagnostics}
\label{section:analysis-group-II}
 
\begin{figure}[tbh]
   \begin{center}
   \includegraphics[width=\textwidth]{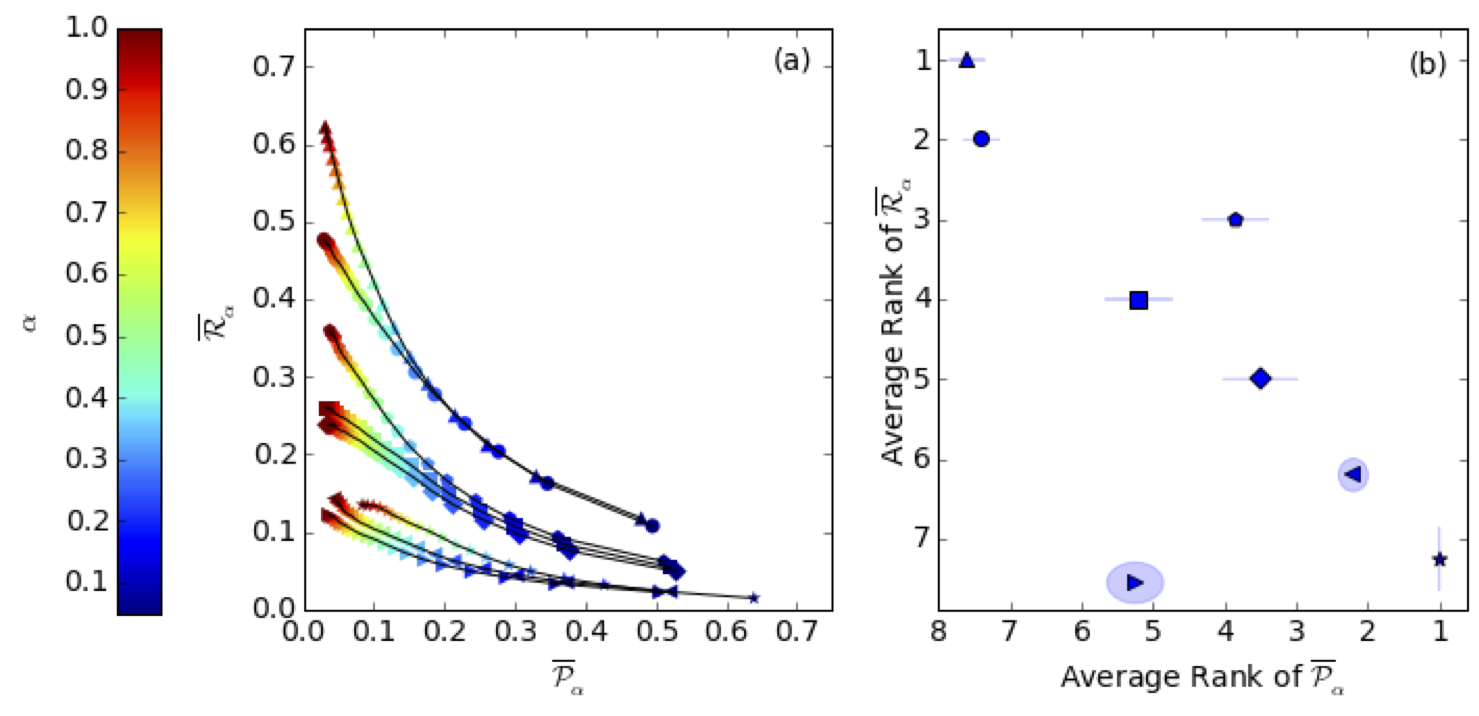}
   \caption{(a)~Average precision-recall curves for the group II pattern diagnostics. (b) Corresponding average precision-recall rank over all Lyapunov vector magnitude thresholds, with an ellipse showing one standard deviation in the ranking distribution in the horizontal and vertical directions. The defects are indicated as follows: topological defect - fast (pentagon), topological defect - appear (right triangle), topological defect - disappear (left triangle), emerging target (star), pinch off (up triangle), 1st derivative (circle), 2nd derivative (square), and 3rd derivative (diamond).  }
    \label{fig:prc-group-II}
    \end{center}
\end{figure}

The precision-recall curves for the group II diagnostics are shown Fig.~\ref{fig:prc-group-II}(a) and the average rankings of the diagnostics are shown in Fig.~\ref{fig:prc-group-II}(b). It is clear that many of these diagnostics perform quite well in comparison with the group I pattern diagnostics shown in Fig.~\ref{fig:prc-group-I}.

It is interesting that the roll pinch-off diagnostic function (up triangles) ranks highest in average recall. For example, the percentage of the area where the Lyapunov magnitude function is 0.95 or larger ($\alpha = 0.95$) that is also covered by the roll pinch-off diagnostic function is just larger than 60\%. These results show that among the group II diagnostics, the roll pinch-off diagnostic is the most effective at locating regions where the Lyapunov magnitude function is above threshold.  This is in agreement with the conclusions drawn from previous work on chaotic convection suggesting the significance of the these defect events in relation to the spatiotemporal dynamics of the leading-order Lyapunov vector~\cite{egolf:2000,scheel:2006,karimi:2012,xu:2018}.  However, the roll pinch-off diagnostic ranks last in precision. This indicates that this diagnostic also selects the most regions, on average, that are not where the Lyapunov magnitude function is large.

The emerging target diagnostic function (stars) ranks first in precision, on average. In fact, the emerging target has the highest precision of all of the group I and II diagnostics. However, the average recall of the emerging target diagnostic quite low and ranks seventh of the group II diagnostics. This indicates that if an emerging target structure is present in the flow field there is a significant probability that the region occupied by the emerging target corresponds to a region where the Lyapunov magnitude function is above threshold. However, since the recall is low this indicates that the emerging target will not be capturing a significant portion of the area where the Lyapunov magnitude function is above threshold. In short, if an emerging target is present it is a good indicator of where the Lyapunov vector is large, on average. 

We next discuss the diagnostic functions that are based upon the dynamics of topological defects. The results from our analysis of the group I diagnostics suggest that the aggregate of all topological defects (shown in Fig.~\ref{fig:prc-group-I}) is a good indicator in terms of both precision and recall. Our intention here is to explore the possibility of increasing the precision and recall performance of a diagnostic function based upon the presence of a topological defect by including some additional measure of its variation with space and time.

Fig.~\ref{fig:prc-group-II} yields that topological defects with a large velocity (pentagons) rank third in average recall and fourth in average precision. These results suggest that topological defects with a large velocity are relatively good indicators of where the Lyapunov magnitude function will be above threshold.

The precision and recall results for the annihilation of topological defects are plotted using left triangles in Fig.~\ref{fig:prc-group-II}.  The annihilation of a topological defect ranks second in precision and sixth in recall. The high value of precision indicates that the annihilation of a topological defect, if present, is a relatively good indicator of where the Lyapunov magnitude function is above threshold. However, since the recall is low the annihilation of a topological defect will not capture a large portion of the domain where the Lyapunov magnitude function is above threshold. Overall, the performance of the annihilation of a topological defect diagnostic function is quite similar to that of the emerging target.

The diagnostic function for the creation of a topological defect is shown in Fig.~\ref{fig:prc-group-II} using right triangles. The precision-recall curves of the creation and annihilation of topological defects shown in Fig.~\ref{fig:prc-group-II}(a) are quite similar. However, there are some interesting differences that are indicated by the average diagnostic results shown in Fig.~\ref{fig:prc-group-II}(b). In particular, the creation of a topological defect performs worse than the annihilation of a topological defect in terms of both recall and precision. In fact, the creation of a topological defect ranks last in recall and sixth in precision among all of the group II diagnostics. We currently do not have an understanding of why the annihilation of topological defect outperforms the creation of a topological defect.

We next discuss the results using temporal derivatives of the temperature field at the horizontal mid-plane as a pattern diagnostic function. We have explored the first order, second order, and third order time derivatives of the temperature field which are represented on Fig.~\ref{fig:prc-group-II} as circles, squares, and diamonds, respectively. The average recall of the first order derivative ranks second, however its precision is poor and ranks seventh. As the order of the derivative increases, the rank of the average recall decreases while the rank of the average precision increases. This indicates that increasing the order of the derivative results in the diagnostic function covering less area where the Lyapunov magnitude function is above threshold while having a higher probability that the area it covers does coincide with regions where the Lyapunov magnitude function is above threshold.

\subsection{Discussion}
\label{section:discussion}

\begin{figure}[tbh]
   \begin{center}
   \includegraphics[width=.65\textwidth]{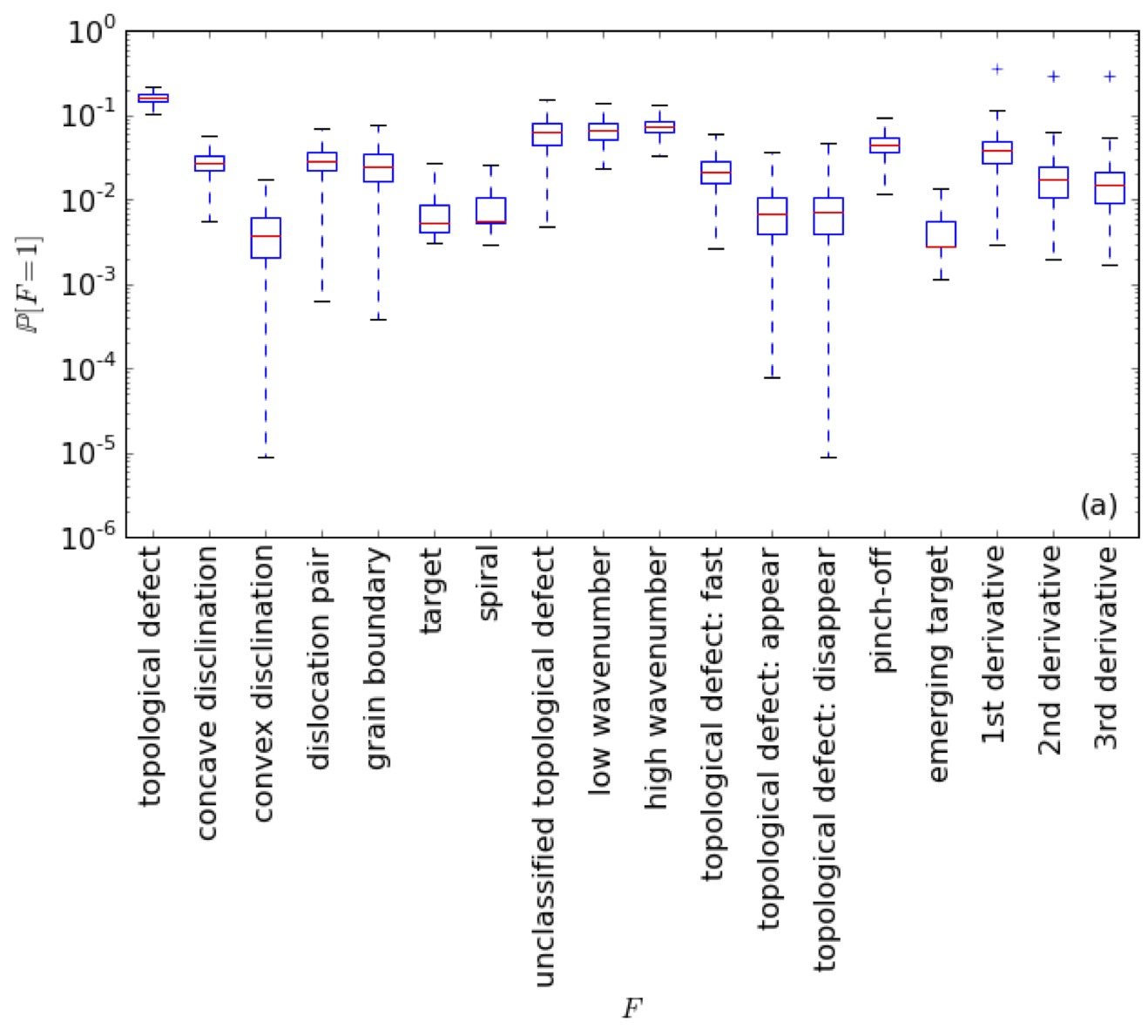} \\
   \includegraphics[width=.65\textwidth]{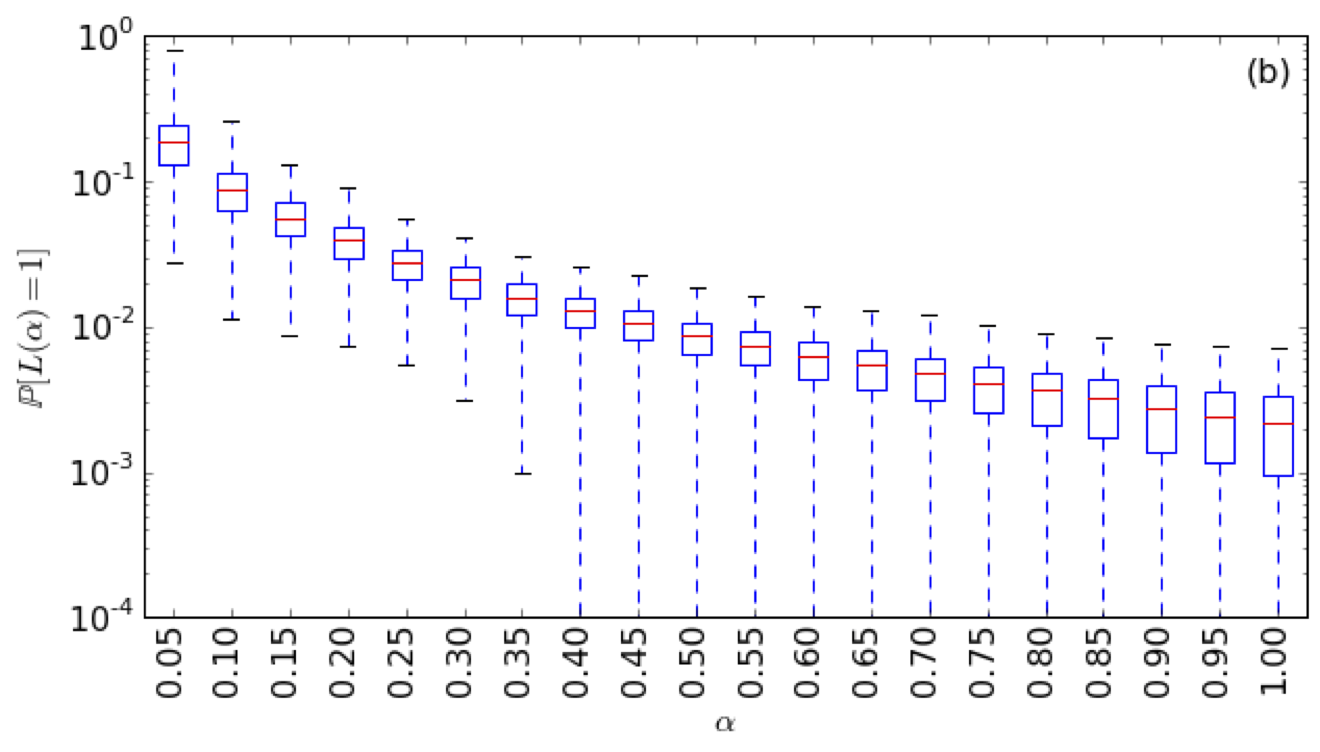}
   \caption{(a)~Box and whisker plot for the nonzero probabilities of each diagnostic of groups I and II. (b)~Box and whisker plot of the probabilities of  the thresholded Lyapunov vector magnitude function.}
    \label{fig:box-and-whisker}
    \end{center}
\end{figure}

In this section we discuss some overall features of the diagnostic features we have explored. Fig.~\ref{fig:box-and-whisker} shows box and whisker plots for the distributions of nonzero probabilities over the entire time series for~(a) the pattern diagnostic regions, and~(b) the thresholded leading-order Lyapunov vector magnitude functions. The red lines in the center of each box denote the median value of the distribution, and the upper and lower extremes of each box give the first and third quartiles, respectively. The whiskers (vertical dashed lines) extend to the maximum and minimum values of the distribution, with the exception of outliers which are indicated by a `+' symbol.

Fig.~\ref{fig:box-and-whisker}(a) shows that the probability distributions for the different pattern diagnostics are, in general, bounded away from one, typically taking up less than ten percent of the domain for the majority of diagnostics. This finding is supported by Figs.~\ref{fig:group-I}-\ref{fig:group-II} which illustrate the regions selected by the different pattern diagnostic functions for a single flow field image. Thus, we can conclude that the recall (vertical axes in Figs.~\ref{fig:prc-group-I}(a) and~\ref{fig:prc-group-II}(a)) is not artificially inflated by our choice of diagnostics. Additionally, the low probability values for Fig.~\ref{fig:box-and-whisker}(b), specifically for larger thresholds, paired with the fairly low probability values in Fig.~\ref{fig:box-and-whisker}(a), indicates that the high recall values in Figs.~\ref{fig:prc-group-I}(a) and~\ref{fig:prc-group-II}(a)) for higher magnitude thresholds are significant. This indicates that the center of the highest peaks are contained within the collection of diagnostics, for if not, the recall would decrease as the Lyapunov vector magnitude threshold increases.

\begin{figure}[tbh]
   \begin{center}
   \includegraphics[width=0.75\textwidth]{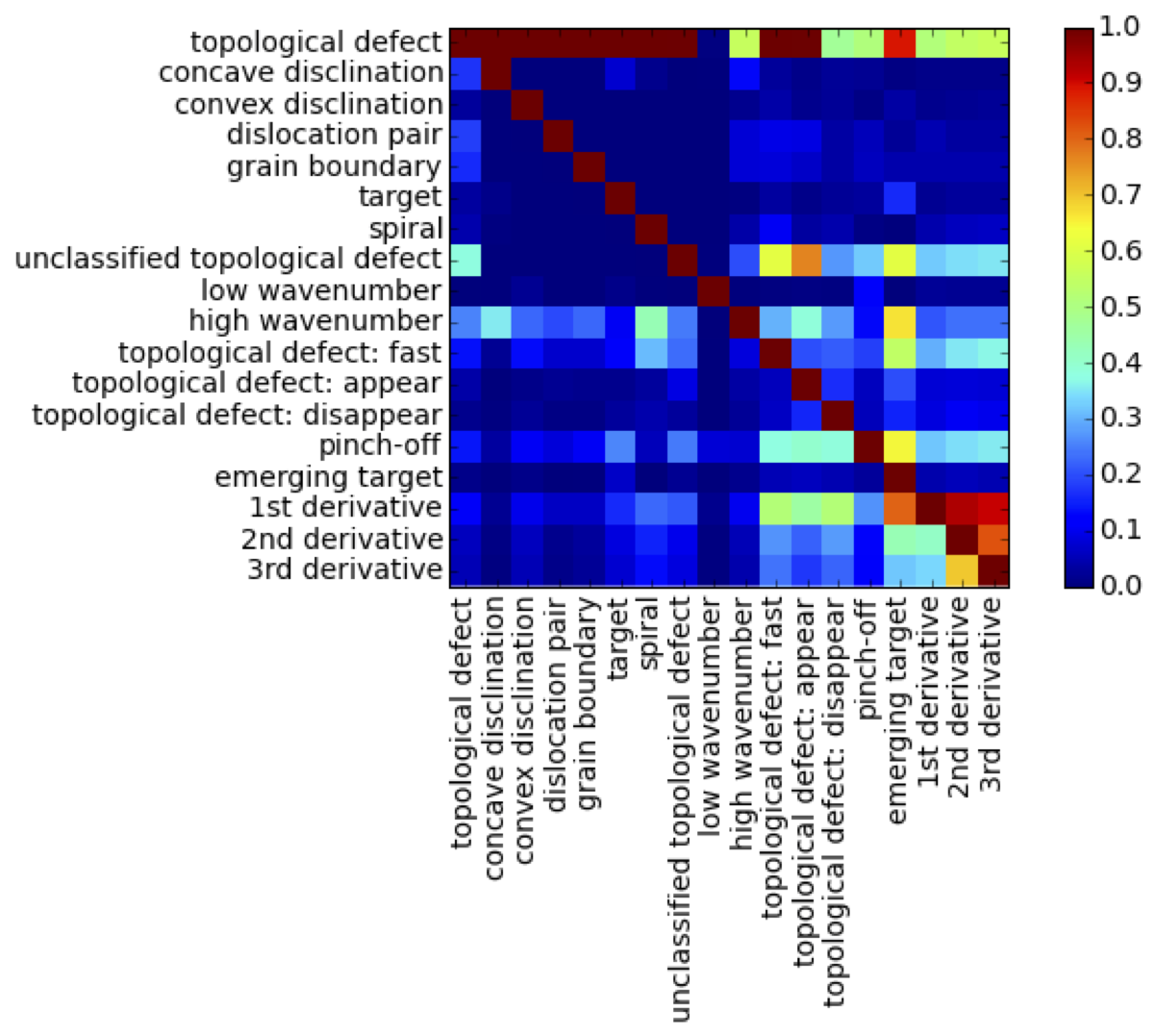}
   \caption{Conditional probability matrix $P_{i,j}$ showing average conditional probabilities where $P_{i,j} = \mathbb{P}(F_i | F_j)$ for every diagnostic $F_i$ we have considered from groups I and II.}
    \label{fig:conditional-probability-matrix}
    \end{center}
\end{figure}

The diagnostics of group I and II were selected using experience and intuition. There are connections between many of them and the diagnostics are not meant to be mutually disjoint with respect to one another. The relationships between the diagnostic functions are illustrated in Fig.~\ref{fig:conditional-probability-matrix} where we plot the conditional probability matrix $P_{i,j}$. The conditional probability matrix is defined as $P_{i,j} = \mathbb{P}(F_i | F_j)$ where $F_i$ and $F_j$ are pattern diagnostic functions and the indices $i$ and $j$ cycle through the 18 different diagnostic functions we have quantified. The magnitude of $P_{i,j}$ is plotted using the color scale shown. Red indicates a large value of $P_{i,j}$ which yields that pattern diagnostics $F_i$ and $F_j$ are related to one another in the sense that their conditional probability is large. Similarly, blue indicates a small value for the conditional probability which yields that the two diagnostics $F_i$ and $F_j$ are not related to one another in a statistical sense. 

Therefore, the diagonal entries where $i=j$ yield $P_{i,i} =1$ since this compares a diagnostic function with itself which are represented as red. The off-diagonal entries indicate the average pairwise conditional probabilities of the different diagnostic functions. The first row of $P_{i,j}$ is for the diagnostic function that is the aggregate of all of the topological defects. The diagnostic functions for the different topological defects from group I are given in rows 2 through 8 as shown in Fig.~\ref{fig:conditional-probability-matrix}. These topological defects are a subset of the diagnostic for the aggregate of all of the topological defects and as a result $P_{1,2:8} = 1$.  It is also clear that the diagnostic for the first order temporal derivative statistically contains the pattern diagnostics for emerging targets and for topological defects with a large velocity. Also, the different orders of the  derivatives are statistically related with one another as expected. In addition, since the specific topological defects and the unclassified topological defects ($F_2, \ldots, F_8$) partition the aggregate diagnostic for all of the topological defects, they are mutually disjoint. This is indicated by the submatrix $P_{2:8, 2:8}$ of Fig.~\ref{fig:conditional-probability-matrix} being approximately a $7 \times 7$ identity matrix. The only notable exception is $P_{2,6}$, which gives the conditional probability of a concave disclination when restricting to target classifications. The reason for this anomaly is described in Appendix A.

\section{Conclusion}
\label{section:conclusion}

Our results suggest a complex relationship between the flow field structures and the leading-order Lyapunov vector of chaotic Rayleigh-B\'enard convection. Our study also shows that ideas from homology can be used to quantify the patterns of chaotic convection. Specifically, we have used ideas from homology to identify spirals, emerging target structures, roll pinch-off events, and to compute the velocity of topological point defects.  Generally speaking, we have found that the group II diagnostics have better precision, but lower recall, than the group I diagnostics. The increased precision is reasonable since the group II diagnostics specifically include aspects of the time evolution of the patterns which were anticipated to be important. 

One aspect that stands out from this study is the high precision of the diagnostic for the emerging target structure. This indicates that if an emerging target is present in the flow field it has the highest probability of all of the pattern diagnostics we have studied here to be in a region where the magnitude of the leading-order Lyapunov vector is above threshold. Also, the pattern diagnostic for the aggregate of all topological point defects outperforms all of the pattern diagnostics in terms of recall.  It is interesting to highlight that we did not find a single type of topological point defect that significantly outperformed the other classified topological defects in terms of precision and recall. This suggests that if one is interested in identifying regions of the chaotic pattern that are  contributing significantly to the chaotic dynamics, a useful first pass of the pattern is to identify the regions containing topological point defects.

For high-dimensional chaotic dynamics there are many Lyapunov vectors that have positive Lyapunov exponents.  For the dynamics we have studied here, there are 24 Lyapunov vectors that have positive Lyapunov exponents. By using only the leading-order Lyapunov vector we are  not using all of the rich information contained by the spectrum of Lyapunov vectors. It is possible that the connection between the patterns and the regions of rapid divergence in the tangent space may become clearer with the inclusion of more Lyapunov vectors.

In our study, we have used the Gram-Schmidt orthonormalized Lyapunov vectors and therefore only the spatiotemporal dynamics of the leading-order Lyapunov vector is physically relevant. However, it is now possible to compute the spectrum of covariant Lyapunov vectors\cite{ginelli:2007,wolfe:2007,pazo:2008,xu:2016,xu:2018} which are not orthonormalized and are expected to contain useful information in their spatiotemporal dynamics.  However, the computation of the covariant Lyapunov vectors in experimentally accessible convection domains is much more expensive and is a topic of future interest.

\section*{Acknowledgements} 

The authors thank the following for many fruitful discussions and insights: Miro Kramar, Jeff Tithof, Balachandra Suri, Brett Tregoning, Logan Kageorge, and Saikat Mukherjee. The authors also thank Shaun Harker for his contributions to the code used to perform this study.

The research was supported by the DARPA MoDyL Program, DARPA contracts HR0011-17-1-0004 and HR0011-16-2-0033, and by NSF grant no. DMS-1622299. The computations were conducted using the resources of the Advanced Research Computing center at Virginia Tech.

\section*{Appendix}
\label{section:appendix}

\subsection{Canonical Defect Classification}

As described in Section~\ref{section:group-I}, topological defects are classified using the following approach.  The topological point defects are clustered using single-linkage hierarchical clustering with a radius of $r = 1$, and then each cluster is classified as follows.

\emph{Concave Disclinations, Convex Disclinations, and Dislocation Pairs.}
Singleton clusters, or monopoles, are classified according to their topological charge.  Defects with positive charge ($+1$) are classified as concave disclinations, and defects with negative charge ($-1$) are classified as convex disclinations.  Any cluster with two topological point defects of opposite charge ($+1, -1$) is classified as a dislocation pair.

\emph{Grain Boundary classification.}
A cluster with three or more points that is arranged in a near-linear sequence of topological point defects of opposite charges is classified as a grain boundary. To detect this situation, we use basic ideas from principal components analysis (PCA). The set of topological point defects is treated as a point cloud in $\mathbb{R}^2$ and then PCA is run on this point cloud. The eigenvalue corresponding to the second principal component is then checked to see if its magnitude falls in the range $\left[\frac{2}{5}r, \frac{3}{2}r\right]$ where $r = 1$, indicating that the points are arranged in a near-linear (but not completely linear) pattern. Finally, the points are orthogonally projected to the first principal component and checked for alternating charges in the resulting ordering. If both of these tests are true, then the cluster is classified as a grain boundary.

\emph{Targets and Spirals.}
Dipole clusters with two positively-charged topological defects are necessarily either targets or spirals. To distinguish the two, we use persistent homology to detect the existence of a local extremum that is positioned between the two topological defects through the method illustrated in Fig.~\ref{fig:defects}.

Targets and spirals are often bounded by two or more concave disclinations, and if the target or spiral is small enough, the clustering radius may group together the target (or spiral) and its surrounding concave disclinations. For example, the target pattern identified in the 6 o'clock position in Fig.~\ref{fig:group-I} is bounded by three concave disclinations. Any cluster of topological defects containing only two positively-charged topological defects, regardless of how many defects are in the cluster, is checked for this scenario as follows. Let $r$ be the radius of the smallest circle circumscribing the two positively-charged topological defects. If the circle of radius $1.25r$ centered at the midpoint between the two positively-charged topological defects does not contain a negatively-charged topological defect, then all negatively-charged topological defects are marked as concave disclinations, and the two positively-charged topological defects are classified using the method described above, as if they formed a dipole.

\vspace{1 em}
\subsection{Matching Algorithm for Topological Defects}

We use a simple algorithm for matching topological defects of like charge from frame to frame in the simulation. Our algorithm rests on the assumption that the vast majority of topological defects are either stationary or move only minimally as the pattern changes, and that the flow is sampled densely enough in time to resolve the majority of any ambiguities. We have found that this approach works well for the chaotic convection data we have explored. 

Let $TD_t(c)$ be the collection of topological defects of charge $c$ at time $t$. For each $x \in TD_t(c)$, we compute the nearest defect $y \in TD_{t+\Delta t}(c)$ and match $x$ to $y$. We then reverse the roles of time points $t$ and $t+\Delta t$ and compute the reverse matchings. Any defects that were matched to each other in both rounds and have distance less than 0.7 from each other are consider matched to each other. All other points are considered  unmatched. The distance filter is included to reduce the chance that a newly annihilated topological defect and a newly created topological defect are not erroneously matched to each other.

\bibstyle{chaos}


\begin{thebibliography}{31}%
\makeatletter
\providecommand \@ifxundefined [1]{%
 \@ifx{#1\undefined}
}%
\providecommand \@ifnum [1]{%
 \ifnum #1\expandafter \@firstoftwo
 \else \expandafter \@secondoftwo
 \fi
}%
\providecommand \@ifx [1]{%
 \ifx #1\expandafter \@firstoftwo
 \else \expandafter \@secondoftwo
 \fi
}%
\providecommand \natexlab [1]{#1}%
\providecommand \enquote  [1]{``#1''}%
\providecommand \bibnamefont  [1]{#1}%
\providecommand \bibfnamefont [1]{#1}%
\providecommand \citenamefont [1]{#1}%
\providecommand \href@noop [0]{\@secondoftwo}%
\providecommand \href [0]{\begingroup \@sanitize@url \@href}%
\providecommand \@href[1]{\@@startlink{#1}\@@href}%
\providecommand \@@href[1]{\endgroup#1\@@endlink}%
\providecommand \@sanitize@url [0]{\catcode `\\12\catcode `\$12\catcode
  `\&12\catcode `\#12\catcode `\^12\catcode `\_12\catcode `\%12\relax}%
\providecommand \@@startlink[1]{}%
\providecommand \@@endlink[0]{}%
\providecommand \url  [0]{\begingroup\@sanitize@url \@url }%
\providecommand \@url [1]{\endgroup\@href {#1}{\urlprefix }}%
\providecommand \urlprefix  [0]{URL }%
\providecommand \Eprint [0]{\href }%
\providecommand \doibase [0]{http://dx.doi.org/}%
\providecommand \selectlanguage [0]{\@gobble}%
\providecommand \bibinfo  [0]{\@secondoftwo}%
\providecommand \bibfield  [0]{\@secondoftwo}%
\providecommand \translation [1]{[#1]}%
\providecommand \BibitemOpen [0]{}%
\providecommand \bibitemStop [0]{}%
\providecommand \bibitemNoStop [0]{.\EOS\space}%
\providecommand \EOS [0]{\spacefactor3000\relax}%
\providecommand \BibitemShut  [1]{\csname bibitem#1\endcsname}%
\let\auto@bib@innerbib\@empty
\bibitem [{\citenamefont {Cross}\ and\ \citenamefont
  {Hohenberg}(1993)}]{cross:1993}%
  \BibitemOpen
  \bibfield  {author} {\bibinfo {author} {\bibfnamefont {M.~C.}\ \bibnamefont
  {Cross}}\ and\ \bibinfo {author} {\bibfnamefont {P.~C.}\ \bibnamefont
  {Hohenberg}},\ }\bibfield  {title} {\enquote {\bibinfo {title} {Pattern
  formation outside of equilibrium},}\ }\href@noop {} {\bibfield  {journal}
  {\bibinfo  {journal} {Rev. Mod. Phys.}\ }\textbf {\bibinfo {volume} {65}},\
  \bibinfo {pages} {851--1112} (\bibinfo {year} {1993})}\BibitemShut {NoStop}%
\bibitem [{\citenamefont {Egolf}\ \emph {et~al.}(2000)\citenamefont {Egolf},
  \citenamefont {Melnikov}, \citenamefont {Pesch},\ and\ \citenamefont
  {Ecke}}]{egolf:2000}%
  \BibitemOpen
  \bibfield  {author} {\bibinfo {author} {\bibfnamefont {D.~A.}\ \bibnamefont
  {Egolf}}, \bibinfo {author} {\bibfnamefont {I.~V.}\ \bibnamefont {Melnikov}},
  \bibinfo {author} {\bibfnamefont {W.}~\bibnamefont {Pesch}}, \ and\ \bibinfo
  {author} {\bibfnamefont {R.~E.}\ \bibnamefont {Ecke}},\ }\bibfield  {title}
  {\enquote {\bibinfo {title} {Mechanisms of extensive spatiotemporal chaos in
  {R}ayleigh-{B}\'{e}nard convection},}\ }\href@noop {} {\bibfield  {journal}
  {\bibinfo  {journal} {Nature}\ }\textbf {\bibinfo {volume} {404}},\ \bibinfo
  {pages} {733--736} (\bibinfo {year} {2000})}\BibitemShut {NoStop}%
\bibitem [{\citenamefont {Duggleby}\ and\ \citenamefont
  {Paul}(2010)}]{duggleby:2010}%
  \BibitemOpen
  \bibfield  {author} {\bibinfo {author} {\bibfnamefont {A.}~\bibnamefont
  {Duggleby}}\ and\ \bibinfo {author} {\bibfnamefont {M.~R.}\ \bibnamefont
  {Paul}},\ }\bibfield  {title} {\enquote {\bibinfo {title} {Computing the
  {K}arhunen-{L}o\`eve dimension of an extensively chaotic flow field given a
  finite amount of data},}\ }\href@noop {} {\bibfield  {journal} {\bibinfo
  {journal} {Computers and Fluids}\ }\textbf {\bibinfo {volume} {39}},\
  \bibinfo {pages} {1704--1710} (\bibinfo {year} {2010})}\BibitemShut {NoStop}%
\bibitem [{\citenamefont {Karimi}\ and\ \citenamefont
  {Paul}(2012)}]{karimi:2012}%
  \BibitemOpen
  \bibfield  {author} {\bibinfo {author} {\bibfnamefont {A.}~\bibnamefont
  {Karimi}}\ and\ \bibinfo {author} {\bibfnamefont {M.~R.}\ \bibnamefont
  {Paul}},\ }\bibfield  {title} {\enquote {\bibinfo {title} {Quantifying
  spatiotemporal chaos in {R}ayleigh-{B}\'enard convection},}\ }\href@noop {}
  {\bibfield  {journal} {\bibinfo  {journal} {Phys. Rev. E}\ }\textbf {\bibinfo
  {volume} {85}},\ \bibinfo {pages} {046201} (\bibinfo {year}
  {2012})}\BibitemShut {NoStop}%
\bibitem [{\citenamefont {Eckmann}\ and\ \citenamefont
  {Ruelle}(1985)}]{eckmann:1985}%
  \BibitemOpen
  \bibfield  {author} {\bibinfo {author} {\bibfnamefont {J.~P.}\ \bibnamefont
  {Eckmann}}\ and\ \bibinfo {author} {\bibfnamefont {D.}~\bibnamefont
  {Ruelle}},\ }\bibfield  {title} {\enquote {\bibinfo {title} {Ergodic theory
  of chaos and strange attractors},}\ }\href@noop {} {\bibfield  {journal}
  {\bibinfo  {journal} {Rev. Mod. Phys.}\ }\textbf {\bibinfo {volume} {57}},\
  \bibinfo {pages} {617--656} (\bibinfo {year} {1985})}\BibitemShut {NoStop}%
\bibitem [{\citenamefont {Wolf}\ \emph {et~al.}(1985)\citenamefont {Wolf},
  \citenamefont {Swift}, \citenamefont {Swinney},\ and\ \citenamefont
  {Vastano}}]{wolf:1985}%
  \BibitemOpen
  \bibfield  {author} {\bibinfo {author} {\bibfnamefont {A.}~\bibnamefont
  {Wolf}}, \bibinfo {author} {\bibfnamefont {J.~B.}\ \bibnamefont {Swift}},
  \bibinfo {author} {\bibfnamefont {H.~L.}\ \bibnamefont {Swinney}}, \ and\
  \bibinfo {author} {\bibfnamefont {A.}~\bibnamefont {Vastano}},\ }\bibfield
  {title} {\enquote {\bibinfo {title} {Determining {L}yapunov exponents from a
  time series},}\ }\href@noop {} {\bibfield  {journal} {\bibinfo  {journal}
  {Physica D}\ }\textbf {\bibinfo {volume} {16}},\ \bibinfo {pages} {285--317}
  (\bibinfo {year} {1985})}\BibitemShut {NoStop}%
\bibitem [{\citenamefont {Bodenschatz}, \citenamefont {Pesch},\ and\
  \citenamefont {Ahlers}(2000)}]{bodenschatz:2000}%
  \BibitemOpen
  \bibfield  {author} {\bibinfo {author} {\bibfnamefont {E.}~\bibnamefont
  {Bodenschatz}}, \bibinfo {author} {\bibfnamefont {W.}~\bibnamefont {Pesch}},
  \ and\ \bibinfo {author} {\bibfnamefont {G.}~\bibnamefont {Ahlers}},\
  }\bibfield  {title} {\enquote {\bibinfo {title} {Recent developments in
  {R}ayleigh-{B}\'enard convection},}\ }\href@noop {} {\bibfield  {journal}
  {\bibinfo  {journal} {Annu. Rev. Fluid Mech.}\ }\textbf {\bibinfo {volume}
  {32}},\ \bibinfo {pages} {709--778} (\bibinfo {year} {2000})}\BibitemShut
  {NoStop}%
\bibitem [{\citenamefont {Paul}\ \emph {et~al.}(2003)\citenamefont {Paul},
  \citenamefont {Chiam}, \citenamefont {Cross}, \citenamefont {Fischer},\ and\
  \citenamefont {Greenside}}]{paul:2003}%
  \BibitemOpen
  \bibfield  {author} {\bibinfo {author} {\bibfnamefont {M.~R.}\ \bibnamefont
  {Paul}}, \bibinfo {author} {\bibfnamefont {K.-H.}\ \bibnamefont {Chiam}},
  \bibinfo {author} {\bibfnamefont {M.~C.}\ \bibnamefont {Cross}}, \bibinfo
  {author} {\bibfnamefont {P.~F.}\ \bibnamefont {Fischer}}, \ and\ \bibinfo
  {author} {\bibfnamefont {H.~S.}\ \bibnamefont {Greenside}},\ }\bibfield
  {title} {\enquote {\bibinfo {title} {Pattern formation and dynamics in
  {R}ayleigh-{B}\'{e}nard convection: numerical simulations of experimentally
  realistic geometries},}\ }\href@noop {} {\bibfield  {journal} {\bibinfo
  {journal} {Physica D}\ }\textbf {\bibinfo {volume} {184}},\ \bibinfo {pages}
  {114--126} (\bibinfo {year} {2003})}\BibitemShut {NoStop}%
\bibitem [{\citenamefont {Scheel}\ and\ \citenamefont
  {Cross}(2006)}]{scheel:2006}%
  \BibitemOpen
  \bibfield  {author} {\bibinfo {author} {\bibfnamefont {J.~D.}\ \bibnamefont
  {Scheel}}\ and\ \bibinfo {author} {\bibfnamefont {M.~C.}\ \bibnamefont
  {Cross}},\ }\bibfield  {title} {\enquote {\bibinfo {title} {Lyapunov
  exponents for small aspect ratio {R}ayleigh-{B}\`{e}nard convection},}\
  }\href@noop {} {\bibfield  {journal} {\bibinfo  {journal} {Phys. Rev. E}\
  }\textbf {\bibinfo {volume} {74}},\ \bibinfo {pages} {066301} (\bibinfo
  {year} {2006})}\BibitemShut {NoStop}%
\bibitem [{\citenamefont {Jayaraman}\ \emph {et~al.}(2006)\citenamefont
  {Jayaraman}, \citenamefont {Scheel}, \citenamefont {Greenside},\ and\
  \citenamefont {Fischer}}]{jayaraman:2006}%
  \BibitemOpen
  \bibfield  {author} {\bibinfo {author} {\bibfnamefont {A.}~\bibnamefont
  {Jayaraman}}, \bibinfo {author} {\bibfnamefont {J.~D.}\ \bibnamefont
  {Scheel}}, \bibinfo {author} {\bibfnamefont {H.~S.}\ \bibnamefont
  {Greenside}}, \ and\ \bibinfo {author} {\bibfnamefont {P.~F.}\ \bibnamefont
  {Fischer}},\ }\bibfield  {title} {\enquote {\bibinfo {title}
  {Characterization of the domain chaos convection state by the largest
  {L}yapunov exponent},}\ }\href@noop {} {\bibfield  {journal} {\bibinfo
  {journal} {Phys. Rev. E}\ }\textbf {\bibinfo {volume} {74}},\ \bibinfo
  {pages} {016209} (\bibinfo {year} {2006})}\BibitemShut {NoStop}%
\bibitem [{\citenamefont {Paul}\ \emph {et~al.}(2007)\citenamefont {Paul},
  \citenamefont {Einarsson}, \citenamefont {Fischer},\ and\ \citenamefont
  {Cross}}]{paul:2007}%
  \BibitemOpen
  \bibfield  {author} {\bibinfo {author} {\bibfnamefont {M.~R.}\ \bibnamefont
  {Paul}}, \bibinfo {author} {\bibfnamefont {M.~I.}\ \bibnamefont {Einarsson}},
  \bibinfo {author} {\bibfnamefont {P.~F.}\ \bibnamefont {Fischer}}, \ and\
  \bibinfo {author} {\bibfnamefont {M.~C.}\ \bibnamefont {Cross}},\ }\bibfield
  {title} {\enquote {\bibinfo {title} {Extensive chaos in
  {R}ayleigh-{B}\'{e}nard convection},}\ }\href@noop {} {\bibfield  {journal}
  {\bibinfo  {journal} {Phys. Rev. E}\ }\textbf {\bibinfo {volume} {75}},\
  \bibinfo {pages} {045203} (\bibinfo {year} {2007})}\BibitemShut {NoStop}%
\bibitem [{\citenamefont {Fischer}(1997)}]{fischer:1997}%
  \BibitemOpen
  \bibfield  {author} {\bibinfo {author} {\bibfnamefont {P.~F.}\ \bibnamefont
  {Fischer}},\ }\bibfield  {title} {\enquote {\bibinfo {title} {An overlapping
  {S}chwarz method for spectral element solution of the incompressible
  {N}avier-{S}tokes equations},}\ }\href@noop {} {\bibfield  {journal}
  {\bibinfo  {journal} {J. Comp. Phys.}\ }\textbf {\bibinfo {volume} {133}},\
  \bibinfo {pages} {84--101} (\bibinfo {year} {1997})}\BibitemShut {NoStop}%
\bibitem [{nek()}]{nek5000}%
  \BibitemOpen
  \href@noop {} {\enquote {\bibinfo {title} {http://nek5000.mcs.anl.gov},}\
  }\BibitemShut {NoStop}%
\bibitem [{\citenamefont {Deville}, \citenamefont {Fischer},\ and\
  \citenamefont {Mund}(2002)}]{deville:2002}%
  \BibitemOpen
  \bibfield  {author} {\bibinfo {author} {\bibfnamefont {M.}~\bibnamefont
  {Deville}}, \bibinfo {author} {\bibfnamefont {P.}~\bibnamefont {Fischer}}, \
  and\ \bibinfo {author} {\bibfnamefont {E.}~\bibnamefont {Mund}},\ }\href@noop
  {} {\emph {\bibinfo {title} {High order methods for incompressible flow}}}\
  (\bibinfo  {publisher} {Cambridge University Press},\ \bibinfo {year}
  {2002})\BibitemShut {NoStop}%
\bibitem [{\citenamefont {Xu}\ and\ \citenamefont {Paul}(2018)}]{xu:2018}%
  \BibitemOpen
  \bibfield  {author} {\bibinfo {author} {\bibfnamefont {M.}~\bibnamefont
  {Xu}}\ and\ \bibinfo {author} {\bibfnamefont {M.~R.}\ \bibnamefont {Paul}},\
  }\bibfield  {title} {\enquote {\bibinfo {title} {Spatiotemporal dynamics of
  the covariant {L}yapunov vectors of chaotic convection},}\ }\href@noop {}
  {\bibfield  {journal} {\bibinfo  {journal} {Phys. Rev. E}\ }\textbf {\bibinfo
  {volume} {97}},\ \bibinfo {pages} {032216} (\bibinfo {year}
  {2018})}\BibitemShut {NoStop}%
\bibitem [{\citenamefont {Kurtuldu}, \citenamefont {Mischaikow},\ and\
  \citenamefont {Schatz}(2011)}]{kurtuldu:2011}%
  \BibitemOpen
  \bibfield  {author} {\bibinfo {author} {\bibfnamefont {H.}~\bibnamefont
  {Kurtuldu}}, \bibinfo {author} {\bibfnamefont {K.}~\bibnamefont
  {Mischaikow}}, \ and\ \bibinfo {author} {\bibfnamefont {M.}~\bibnamefont
  {Schatz}},\ }\bibfield  {title} {\enquote {\bibinfo {title} {Extensive
  scaling from computational homology and {K}arhunen-{L}\`oeve decomposition of
  {R}ayleigh-{B}\'enard convection experiments},}\ }\href@noop {} {\bibfield
  {journal} {\bibinfo  {journal} {Phys. Rev. Lett.}\ }\textbf {\bibinfo
  {volume} {107}},\ \bibinfo {pages} {034503} (\bibinfo {year}
  {2011})}\BibitemShut {NoStop}%
\bibitem [{\citenamefont {Kram\'ar}\ \emph {et~al.}(2016)\citenamefont
  {Kram\'ar}, \citenamefont {Levanger}, \citenamefont {Tithof}, \citenamefont
  {Suri}, \citenamefont {Xu}, \citenamefont {Paul}, \citenamefont {Schatz},\
  and\ \citenamefont {Mischaikow}}]{kramar:2016}%
  \BibitemOpen
  \bibfield  {author} {\bibinfo {author} {\bibfnamefont {K.}~\bibnamefont
  {Kram\'ar}}, \bibinfo {author} {\bibfnamefont {R.}~\bibnamefont {Levanger}},
  \bibinfo {author} {\bibfnamefont {J.}~\bibnamefont {Tithof}}, \bibinfo
  {author} {\bibfnamefont {B.}~\bibnamefont {Suri}}, \bibinfo {author}
  {\bibfnamefont {M.}~\bibnamefont {Xu}}, \bibinfo {author} {\bibfnamefont
  {M.~R.}\ \bibnamefont {Paul}}, \bibinfo {author} {\bibfnamefont {M.~F.}\
  \bibnamefont {Schatz}}, \ and\ \bibinfo {author} {\bibfnamefont
  {K.}~\bibnamefont {Mischaikow}},\ }\bibfield  {title} {\enquote {\bibinfo
  {title} {Analysis of {K}olmogorov flow and {R}ayleigh-{B}\'enard convection
  using persistent homology},}\ }\href@noop {} {\bibfield  {journal} {\bibinfo
  {journal} {Physica D}\ }\textbf {\bibinfo {volume} {334}},\ \bibinfo {pages}
  {82--98} (\bibinfo {year} {2016})}\BibitemShut {NoStop}%
\bibitem [{\citenamefont {Cross}\ and\ \citenamefont
  {Newell}(1984)}]{cross:1984}%
  \BibitemOpen
  \bibfield  {author} {\bibinfo {author} {\bibfnamefont {M.~C.}\ \bibnamefont
  {Cross}}\ and\ \bibinfo {author} {\bibfnamefont {A.~C.}\ \bibnamefont
  {Newell}},\ }\bibfield  {title} {\enquote {\bibinfo {title} {Convection
  patterns in large aspect ratio systems},}\ }\href@noop {} {\bibfield
  {journal} {\bibinfo  {journal} {Physica D}\ }\textbf {\bibinfo {volume}
  {10}},\ \bibinfo {pages} {299--328} (\bibinfo {year} {1984})}\BibitemShut
  {NoStop}%
\bibitem [{\citenamefont {Kaplan}\ and\ \citenamefont
  {Yorke}(1979)}]{kaplan:1979}%
  \BibitemOpen
  \bibfield  {author} {\bibinfo {author} {\bibfnamefont {J.~L.}\ \bibnamefont
  {Kaplan}}\ and\ \bibinfo {author} {\bibfnamefont {J.~A.}\ \bibnamefont
  {Yorke}},\ }\bibfield  {title} {\enquote {\bibinfo {title} {Lecture notes in
  math},}\ }\href@noop {} {\bibfield  {journal} {\bibinfo  {journal} {730}\ ,\
  \bibinfo {pages} {204}} (\bibinfo {year} {1979})}\BibitemShut {NoStop}%
\bibitem [{\citenamefont {Saito}\ and\ \citenamefont
  {Rehmsmeier}(2015)}]{Saito:2015}%
  \BibitemOpen
  \bibfield  {author} {\bibinfo {author} {\bibfnamefont {T.}~\bibnamefont
  {Saito}}\ and\ \bibinfo {author} {\bibfnamefont {M.}~\bibnamefont
  {Rehmsmeier}},\ }\bibfield  {title} {\enquote {\bibinfo {title} {The
  precision-recall plot is more informative than the roc plot when evaluating
  binary classifiers on imbalanced datasets},}\ }\href {\doibase
  10.1371/journal.pone.0118432} {\bibfield  {journal} {\bibinfo  {journal}
  {PLOS ONE}\ }\textbf {\bibinfo {volume} {10}},\ \bibinfo {pages} {1--21}
  (\bibinfo {year} {2015})}\BibitemShut {NoStop}%
\bibitem [{\citenamefont {He}\ and\ \citenamefont {Garcia}(2009)}]{He:2009}%
  \BibitemOpen
  \bibfield  {author} {\bibinfo {author} {\bibfnamefont {H.}~\bibnamefont
  {He}}\ and\ \bibinfo {author} {\bibfnamefont {E.~A.}\ \bibnamefont
  {Garcia}},\ }\bibfield  {title} {\enquote {\bibinfo {title} {Learning from
  imbalanced data},}\ }\href {\doibase 10.1109/TKDE.2008.239} {\bibfield
  {journal} {\bibinfo  {journal} {IEEE Transactions on Knowledge and Data
  Engineering}\ }\textbf {\bibinfo {volume} {21}},\ \bibinfo {pages}
  {1263--1284} (\bibinfo {year} {2009})}\BibitemShut {NoStop}%
\bibitem [{\citenamefont {Soille}(2013)}]{soille:2013}%
  \BibitemOpen
  \bibfield  {author} {\bibinfo {author} {\bibfnamefont {P.}~\bibnamefont
  {Soille}},\ }\href {https://books.google.com/books?id=ZFzxCAAAQBAJ} {\emph
  {\bibinfo {title} {Morphological Image Analysis: Principles and
  Applications}}}\ (\bibinfo  {publisher} {Springer Berlin Heidelberg},\
  \bibinfo {year} {2013})\BibitemShut {NoStop}%
\bibitem [{\citenamefont {Chaikin}\ and\ \citenamefont
  {Lubensky}(2000)}]{chaikin:2000}%
  \BibitemOpen
  \bibfield  {author} {\bibinfo {author} {\bibfnamefont {P.~M.}\ \bibnamefont
  {Chaikin}}\ and\ \bibinfo {author} {\bibfnamefont {T.~C.}\ \bibnamefont
  {Lubensky}},\ }\href@noop {} {\emph {\bibinfo {title} {Principles of
  Condensed Matter Physics}}}\ (\bibinfo  {publisher} {Cambridge University
  Press},\ \bibinfo {year} {2000})\BibitemShut {NoStop}%
\bibitem [{\citenamefont {Bazen}\ and\ \citenamefont
  {Gerez}(2002)}]{bazen:2002}%
  \BibitemOpen
  \bibfield  {author} {\bibinfo {author} {\bibfnamefont {A.~M.}\ \bibnamefont
  {Bazen}}\ and\ \bibinfo {author} {\bibfnamefont {S.~H.}\ \bibnamefont
  {Gerez}},\ }\bibfield  {title} {\enquote {\bibinfo {title} {Systematic
  methods for the computation of the directional fields and singular points of
  fingerprints},}\ }\href {\doibase 10.1109/TPAMI.2002.1017618} {\bibfield
  {journal} {\bibinfo  {journal} {IEEE Trans. Pattern Anal. Mach. Intell.}\
  }\textbf {\bibinfo {volume} {24}},\ \bibinfo {pages} {905--919} (\bibinfo
  {year} {2002})}\BibitemShut {NoStop}%
\bibitem [{\citenamefont {Egolf}, \citenamefont {Melnikov},\ and\ \citenamefont
  {Bodenschatz}(1998)}]{egolf:1998}%
  \BibitemOpen
  \bibfield  {author} {\bibinfo {author} {\bibfnamefont {D.~A.}\ \bibnamefont
  {Egolf}}, \bibinfo {author} {\bibfnamefont {I.~V.}\ \bibnamefont {Melnikov}},
  \ and\ \bibinfo {author} {\bibfnamefont {E.}~\bibnamefont {Bodenschatz}},\
  }\bibfield  {title} {\enquote {\bibinfo {title} {Importance of local pattern
  properties in spiral defect chaos},}\ }\href {\doibase
  10.1103/PhysRevLett.80.3228} {\bibfield  {journal} {\bibinfo  {journal}
  {Phys. Rev. Lett.}\ }\textbf {\bibinfo {volume} {80}},\ \bibinfo {pages}
  {3228--3231} (\bibinfo {year} {1998})}\BibitemShut {NoStop}%
\bibitem [{\citenamefont {Becker}\ and\ \citenamefont
  {Ahlers}(2006)}]{becker:2006}%
  \BibitemOpen
  \bibfield  {author} {\bibinfo {author} {\bibfnamefont {N.}~\bibnamefont
  {Becker}}\ and\ \bibinfo {author} {\bibfnamefont {G.}~\bibnamefont
  {Ahlers}},\ }\bibfield  {title} {\enquote {\bibinfo {title} {Local wave
  director analysis of domain chaos in {R}ayleigh-{B}\'{e}nard convection},}\
  }\href {http://stacks.iop.org/1742-5468/2006/i=12/a=P12002} {\bibfield
  {journal} {\bibinfo  {journal} {Journal of Statistical Mechanics: Theory and
  Experiment}\ }\textbf {\bibinfo {volume} {2006}},\ \bibinfo {pages} {P12002}
  (\bibinfo {year} {2006})}\BibitemShut {NoStop}%
\bibitem [{\citenamefont {Busse}(1978)}]{busse:1978}%
  \BibitemOpen
  \bibfield  {author} {\bibinfo {author} {\bibfnamefont {F.~H.}\ \bibnamefont
  {Busse}},\ }\bibfield  {title} {\enquote {\bibinfo {title} {Non-linear
  properties of thermal convection},}\ }\href@noop {} {\bibfield  {journal}
  {\bibinfo  {journal} {Reports on Prog. Phys.}\ }\textbf {\bibinfo {volume}
  {41}},\ \bibinfo {pages} {1929--1967} (\bibinfo {year} {1978})}\BibitemShut
  {NoStop}%
\bibitem [{\citenamefont {Ginelli}\ \emph {et~al.}(2007)\citenamefont
  {Ginelli}, \citenamefont {Poggi}, \citenamefont {Turchi}, \citenamefont
  {Chat\`{e}}, \citenamefont {Livi},\ and\ \citenamefont
  {Politi}}]{ginelli:2007}%
  \BibitemOpen
  \bibfield  {author} {\bibinfo {author} {\bibfnamefont {F.}~\bibnamefont
  {Ginelli}}, \bibinfo {author} {\bibfnamefont {P.}~\bibnamefont {Poggi}},
  \bibinfo {author} {\bibfnamefont {A.}~\bibnamefont {Turchi}}, \bibinfo
  {author} {\bibfnamefont {H.}~\bibnamefont {Chat\`{e}}}, \bibinfo {author}
  {\bibfnamefont {R.}~\bibnamefont {Livi}}, \ and\ \bibinfo {author}
  {\bibfnamefont {A.}~\bibnamefont {Politi}},\ }\bibfield  {title} {\enquote
  {\bibinfo {title} {Characterizing dynamics with covariant {L}yapunov
  vectors},}\ }\href@noop {} {\bibfield  {journal} {\bibinfo  {journal} {Phys.
  Rev. Lett.}\ }\textbf {\bibinfo {volume} {99}},\ \bibinfo {pages} {130601}
  (\bibinfo {year} {2007})}\BibitemShut {NoStop}%
\bibitem [{\citenamefont {Wolfe}\ and\ \citenamefont
  {Samelson}(2007)}]{wolfe:2007}%
  \BibitemOpen
  \bibfield  {author} {\bibinfo {author} {\bibfnamefont {C.~L.}\ \bibnamefont
  {Wolfe}}\ and\ \bibinfo {author} {\bibfnamefont {R.~M.}\ \bibnamefont
  {Samelson}},\ }\bibfield  {title} {\enquote {\bibinfo {title} {An efficient
  method for recovering {L}yapunov vectors from singular vectors},}\
  }\href@noop {} {\bibfield  {journal} {\bibinfo  {journal} {Tellus}\ }\textbf
  {\bibinfo {volume} {59A}},\ \bibinfo {pages} {355--366} (\bibinfo {year}
  {2007})}\BibitemShut {NoStop}%
\bibitem [{\citenamefont {Paz\'o}\ \emph {et~al.}(2008)\citenamefont {Paz\'o},
  \citenamefont {Szendro}, \citenamefont {L\'opez},\ and\ \citenamefont
  {Rodr\'iguez}}]{pazo:2008}%
  \BibitemOpen
  \bibfield  {author} {\bibinfo {author} {\bibfnamefont {D.}~\bibnamefont
  {Paz\'o}}, \bibinfo {author} {\bibfnamefont {I.~G.}\ \bibnamefont {Szendro}},
  \bibinfo {author} {\bibfnamefont {J.~M.}\ \bibnamefont {L\'opez}}, \ and\
  \bibinfo {author} {\bibfnamefont {M.~A.}\ \bibnamefont {Rodr\'iguez}},\
  }\bibfield  {title} {\enquote {\bibinfo {title} {Structure of characteristic
  {L}yapunov vectors in spatiotemporal chaos},}\ }\href@noop {} {\bibfield
  {journal} {\bibinfo  {journal} {Phys. Rev. E}\ }\textbf {\bibinfo {volume}
  {78}},\ \bibinfo {pages} {016209} (\bibinfo {year} {2008})}\BibitemShut
  {NoStop}%
\bibitem [{\citenamefont {Xu}\ and\ \citenamefont {Paul}(2016)}]{xu:2016}%
  \BibitemOpen
  \bibfield  {author} {\bibinfo {author} {\bibfnamefont {M.}~\bibnamefont
  {Xu}}\ and\ \bibinfo {author} {\bibfnamefont {M.~R.}\ \bibnamefont {Paul}},\
  }\bibfield  {title} {\enquote {\bibinfo {title} {Covariant {L}yapunov vectors
  of chaotic {R}ayleigh-{B}\'enard convection},}\ }\href@noop {} {\bibfield
  {journal} {\bibinfo  {journal} {Phys. Rev. E}\ }\textbf {\bibinfo {volume}
  {93}},\ \bibinfo {pages} {062208} (\bibinfo {year} {2016})}\BibitemShut
  {NoStop}%
\end{thebibliography}

%

\end{document}